\documentclass[apj]{emulateapj}
\submitted{Submitted to The Astrophysical Journal on 2011, June 6; accepted for publication on 2011, September 11. }

\usepackage{color, rotating}

\def\Ha{\ifmmode^{\mathrm{H}\alpha }\else$\mathrm{H}\alpha$\fi}
\def\Hb{\ifmmode^{\mathrm{H}\beta }\else$\mathrm{H}\beta$\fi}
\def\LyA{\ifmmode^{\mathrm{H}\alpha }\else$\mathrm{Ly}\alpha$\fi}
\def\BrA{\ifmmode^{\mathrm{Br}\alpha }\else$\mathrm{Br}\alpha$\fi}
\def\BrG{\ifmmode^{\mathrm{Br}\gamma }\else$\mathrm{Br}\gamma$\fi}
\def\PaB{\ifmmode^{\mathrm{Pa}\beta }\else$\mathrm{Pa}\beta$\fi}
\def\mag{\ifmmode^{\rm m }\else$^{\rm m}$\fi}

\def\as{$\,^{\prime\prime}\,$}

\def\hh{\ifmmode^{\rm h}\else$^{\rm h}$\fi}
\def\mm{\ifmmode^{\rm m}\else$^{\rm m}$\fi}
\def\ss{\ifmmode^{\rm s}\else$^{\rm s}$\fi}
\def\deg{\ifmmode^\circ\else$^\circ $\fi}
\def\amin{\ifmmode^\prime\else$^\prime $\fi}

\def\decdm#1#2{\ifmmode{#1}\else{$#1$}\fi\deg\ #2\amin\ }

\def\dec#1#2#3{\ifmmode{#1}\else{$#1$}\fi\deg\ #2\amin\ #3\as\ }

\def\decb#1#2#3#4{\ifmmode{#1}\else{$#1$}\fi\deg\ #2\amin\ #3\farcs#4 }

\shorttitle{Gas distribution, kinematics, and excitation structure in the disks around $\beta$~CMi and $\zeta$~Tau}
\shortauthors{}

\begin{document}


\title{Gas distribution, kinematics, and excitation structure in the disks around the classical Be stars $\beta$~Canis~Minoris and $\zeta$~Tauri \footnotemark[1]}

\footnotetext[1]{Based on observations made with ESO telescopes 
at the Paranal Observatory under programme IDs 
084.C-0848(A) and 085.C-0911(A) and with the CHARA array.
}


\author{S.~Kraus\altaffilmark{1}, 
  J.D.~Monnier\altaffilmark{1}, 
  X.~Che\altaffilmark{1}, 
  G.~Schaefer\altaffilmark{2}, 
  Y.~Touhami\altaffilmark{3},
  D.R.~Gies\altaffilmark{3},
  J.P.~Aufdenberg\altaffilmark{4},
  F.~Baron\altaffilmark{1},
  N.~Thureau\altaffilmark{5},
  T.A.~ten Brummelaar\altaffilmark{2},
  H.A.~McAlister\altaffilmark{2},
  N.H.~Turner\altaffilmark{2},
  J.~Sturmann\altaffilmark{2},
  L.~Sturmann\altaffilmark{2}
}

\email{stefankr@umich.edu}

\affil{
$^{1}$~Department of Astronomy, University of Michigan, 918 Dennison Building, Ann Arbor, MI 48109-1090, USA\\
$^{2}$~The CHARA Array, Georgia State University, P.O.\ Box 3965, Atlanta, GA 30302-3965, USA\\
$^{3}$~Center for High Angular Resolution Astronomy and Department of Physics and Astronomy, Georgia State University, P.O.\ Box 4106, Atlanta, GA 30302-4106, USA\\
$^{4}$~Department of Physical Sciences, Embry-Riddle Aeronautical University, 600 S.\ Clyde Morris Blvd., Daytona Beach FL 32114, USA\\
$^{5}$~Department of Physics and Astronomy, University of St. Andrews, Scotland, UK
}


\begin{abstract}
Using CHARA and VLTI near-infrared spectro-interferometry with hectometric baseline lengths (up to 330\,m)
and with high spectral resolution (up to $\lambda/\Delta\lambda=12\,000$),
we studied the gas distribution and kinematics around two classical Be stars.
The combination of high spatial and spectral resolution achieved allows us to constrain
the gas velocity field on scales of a few stellar radii and 
to obtain, for the first time in optical interferometry,
a dynamical mass estimate using the position-velocity analysis technique known from radio astronomy.
For our first target star, $\beta$~Canis~Minoris, we model the H+K-band continuum and
Br$\gamma$-line geometry with a near-critical rotating stellar photosphere and 
a geometrically thin equatorial disk.
Testing different disk rotation laws, we find that the disk is in
Keplerian rotation ($v(r) \propto r^{-0.5\pm0.1}$) and derive the disk position angle ($140 \pm 1.7^{\circ}$)
inclination ($38.5\pm1^{\circ}$), and the mass of the central star ($3.5\pm0.2~M_{\sun}$).
As a second target star, we observed the prototypical Be star $\zeta$~Tauri and 
spatially resolved the Br$\gamma$ emission as well as nine transitions from the hydrogen Pfund series (Pf\,14-22).
Comparing the spatial origin of the different line transitions, we find that the Brackett (Br$\gamma$), 
Pfund (Pf\,14-17), and Balmer (H$\alpha$) lines originate from different stellocentric radii 
($R_{\rm cont}<R_{\rm Pf}<R_{{\rm Br}\gamma}\sim R_{\rm H\alpha}$), 
which we can reproduce with an LTE line radiative transfer computation.
Discussing different disk-formation scenarios, we conclude that our constraints 
are inconsistent with wind compression models predicting a strong outflowing velocity component,
but support viscous decretion disk models, where the Keplerian-rotating disk is
replenished with material from the near-critical rotating star.
\end{abstract}





\section{Introduction}

Classical Be stars are main-sequence (or near main-sequence) B-type stars 
associated with hydrogen line emission, indicating the presence of ionized
circumstellar gas, which is believed to be arranged in an equatorial disk-like structure.
Optical/infrared spectroscopic observations have revealed
non-radial pulsations \citep[e.g.][]{riv98}, which might provide a way to feed material from the photosphere
to the inner disk. Polarimetric studies \citep[e.g.][]{dra11} can constrain
the disk density structure.
Further unique insights into the structure and physics of these disks can also
be obtained with interferometry at visual and infrared wavelengths, allowing
one to unravel the inner disk structure on scales of a few stellar radii directly.
For instance, interferometric studies have allowed
associating quasi-cyclic variations in the ratio between the blue- and red-shifted wing 
of the H$\alpha$-line emission ($V/R$ variability) with global oscillations
in the circumstellar disk, likely in the form of a one-armed spiral density pattern
\citep[e.g.][]{vak98,ste09,car09,sch10}.
Besides studies on the disk continuum geometry, interferometric observations in spectral
lines have provided the first direct constraints on the gas kinematics,
in particular for the hydrogen spectral lines of the
Balmer \citep[e.g.][]{qui94,vak98,tyc05,del11}, Brackett \citep{mei07,mei11}, and 
Pfund series \citep{pot10}.
These studies have provided growing evidence that the disks around 
classical Be stars exhibit a near-Keplerian rotation profile, which might allow
to effectively rule out several disk-formation scenarios 
(e.g.\ see review by \citealt{car10}).
However, most earlier studies using spectro-interferometry were 
limited in terms of spectral resolution or baseline position angle (PA) coverage,
leaving significant uncertainties about the detailed gas velocity field
and the evidence to distinguish between a purely rotational versus 
an expanding velocity component in the disk.
Obtaining such evidence is essential in order to decide between different 
scenarios which have been proposed to explain the disk formation mechanism, including 
radiatively driven winds, ram pressure or magnetically induced wind compression, 
and viscous decretion \citep{por03}.
Furthermore, recently there has been a controversy about the
appearance of a phase inversion in spectro-interferometric observations 
of several classical Be stars, which triggered speculations 
about secondary dynamical effects or the need for an additional kinematical component 
beyond the canonical star+disk paradigm \citep{ste11}.

Here, we present near-infrared spectro-interferometric observations on
the classical Be star $\beta$\,CMi with a high spectral resolution 
of $R=12,000$ in the hydrogen Br$\gamma$-line,
enabling us to constrain the rotation profile directly.
In addition, we observed the classical Be star $\zeta$\,Tau for the first time
in multiple hydrogen line transitions (Br$\gamma$ and Pfund lines),
providing direct information about the excitation structure within the disk.

In the following, we present our CHARA and VLTI interferometric 
observations (Sect.~\ref{sec:observations}).
The observations in spectral lines are then first interpreted using
a model-independent photocenter analysis approach (Sect.~\ref{sec:photocenter}).
In Sect.~\ref{sec:bcmi}, we present our continuum and kinematical modeling on $\beta$\,CMi,
followed by our discussion of the results on $\zeta$\,Tau (Sect.~\ref{sec:ztau}).
Finally, we summarize our findings in Sect.~\ref{sec:conclusions}.

\section{Observations}
\label{sec:observations}

\begin{sidewaystable}
\begin{center}
\caption{Observation log of our CHARA and VLTI observations.}
\label{tab:obslog}
\begin{tabular}{cccccccc}
  \tableline\tableline
  \noalign{\smallskip}
  Target       & Date        & Instrument & Spectral   & NP       & DIT      & Telescope     &  Calibrator(s) \\
               & (UT)        &            & mode       &          & [s]      & configuration &             \\
  \noalign{\smallskip}
  \tableline
  \noalign{\smallskip}
  $\beta$\,CMi & 2009-12-31 & AMBER       & HR-K~2.172 & 1        & 1        & UT1-UT2-UT4 & HD\,71095 \\
               & 2010-04-23 & AMBER       & HR-K~2.172 & 1        & 6        & K0-G1-A0    & HD\,71095 \\
               & 2008-12-02 & MIRC        & H35        & 1        & -        & S1-E1-W1-W2 & HD\,25490, HD\,79469, HD\,97633 \\
               & 2009-11-10 & MIRC        & H35        & 3        & -        & S1-E1-W1-W2 & HD\,43042 \\
               & 2010-11-02 & MIRC        & H35        & 3        & -        & S1-E1-W1-W2 & HD\,43042 \\
               & 2010-11-03 & MIRC        & H35        & 2        & -        & S2-E2-W1-W1 & HD\,43042 \\
               & 2010-12-14 & MIRC        & H35        & 1        & -        & S1-E2-W1-W2 & HD\,79469, HD\,97633 \\
               & 2010-11-30 & CLIMB       & K          & 2        & -        & S2-W1-W2    & HD\,73262 \\
               & 2010-12-02 & CLIMB       & K          & 1        & -        & S1-E2-W1    & HD\,73262 \\
               & 2010-12-03 & CLIMB       & K          & 4        & -        & S2-E2-W2    & HD\,73262 \\
  \noalign{\smallskip}
  \tableline
  \noalign{\smallskip}
  $\zeta$\,Tau & 2010-01-01 & AMBER       & MR-K~2.3   & 1        & 0.2      & UT1-UT2-UT4 & HD\,71095 \\
  \noalign{\smallskip}
  \tableline
\end{tabular}
\tablecomments{
  Column~5 (NP) denotes the number of pointings obtained on the science star.
  For the calibrators, we assume the following uniform disk (UD) diameters:
  For
  HD\,25490 ($0.536 \pm 0.037$~mas),
  HD\,43042 ($0.62 \pm 0.05$~mas),
  HD\,71095 ($2.009 \pm 0.139$~mas), and 
  HD\,73262 ($0.47 \pm 0.03$~mas) we use the
  diameter estimates from searchCal \citep{bon06}.
  The UD diameters of 
  HD\,79469 ($0.471 \pm 0.027$~mas) and
  HD\,97633 ($0.68 \pm 0.06$~mas)
  have been estimated by averaging the estimates
  from three independent photometric methods
  \citep{bar78,bon06,ker08}. 
}
\end{center}
\end{sidewaystable}

\begin{figure}[bthp]
  \centering
  \includegraphics[angle=0,scale=0.4]{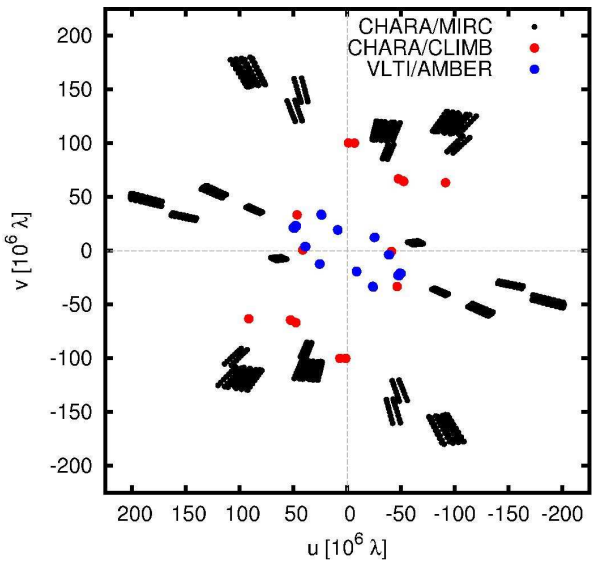}
  \caption{
    \footnotesize
    $uv$-coverage achieved with our CHARA/MIRC ($H$-band),
    CHARA/CLIMB ($K$-band), 
    and VLTI/AMBER ($K$-band) interferometric observations 
    on $\beta$\,CMi.
  }
  \label{fig:uvcov}
\end{figure}

\begin{figure}[t]
  \centering
  \includegraphics[angle=0,scale=0.65]{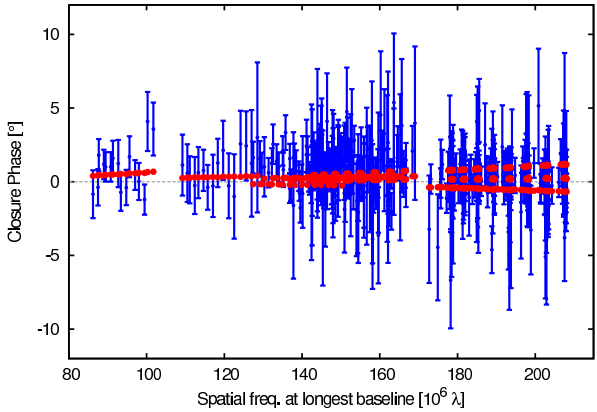}
  \caption{
    \footnotesize
    Closure phases measured with the MIRC beam combiner 
    on $\beta$\,CMi (blue data points), overplotted with the
    model predictions from our best-fit $H$-band model (red data points; Sect.~\ref{sec:bcmicont}).
  }
  \label{fig:bcmicp}
\end{figure}

Our near-infrared $H$-band continuum observations on $\beta$\,CMi were obtained
using the CHARA array \citep{ten05}, which is operated by Georgia State University.
The MIRC beam combiner \citep{mon06b} allowed us to combine the light 
from four of the six CHARA 1\,m-telescopes simultaneously, yielding baseline lengths of 
up to 330\,m, with good baseline coverage (Fig.~\ref{fig:uvcov}). 
The MIRC data cover the $H$-band with low spectral dispersion ($R=35$)
and was reduced using the University of Michigan MIRC data reduction pipeline \citep{mon07}.

In order to investigate the $K$-band disk geometry of $\beta$\,CMi, we employed
the CHARA/CLIMB 3-telescope beam combiner \citep{stu10}.
Visibilities and closure phases were derived using the
``redclimb'' and ``reduceir'' software.  Besides the statistical errors, 
we also add a calibration uncertainty of 0.05 for the derived visibilities, 
which represents an empirical value for the typical scatter 
in the instrument transfer function.

Spectro-interferometric observations with medium (MR mode, $R=1500$) and high spectral dispersion (HR mode, $R=12\,000$) 
were obtained with the \textit{Very Large Telescope Interferometer} (VLTI) of the
European Southern Observatory and the AMBER 3-telescope beam combiner instrument \citep{pet07}.
The AMBER observations on $\beta$\,CMi were recorded using
three 8.2\,m unit telescopes (2009-12-31) and 
three 1.8\,m auxiliary telescopes (2010-04-23), respectively.
The atmospheric piston was stabilized using the FINITO fringe tracker \citep{leb08}, 
which allowed us to use long detector integration times (DITs) of 1\,s and 6\,s
and to record data with high spectral dispersion 
around the hydrogen Br$\gamma$-line ($\lambda_{\rm Br\gamma}^{\rm vacuum}=2.166078~\mu$m).
Due to the presence of residual phase jitter, the absolute visibility calibration 
of the AMBER data is not reliable, while the important 
wavelength-differential observables are not affected.
Spectra and wavelength-differential visibilities and phases (DPs, Fig.~\ref{fig:bcmiphotocenter}) 
were derived from the AMBER data using the amdlib (V3.0) data reduction software \citep{tat07b,che09}.
The wavelength calibration was done using atmospheric telluric features 
close to the Br$\gamma$-line (yielding an accuracy of approximately 1~spectral channel) 
and by applying a heliocentric-barycentric system correction using
heliocentric velocities of $+5.97$~km\,s$^{-1}$ (2009-12-31) 
and $-28.61$~km\,s$^{-1}$ (2010-04-23), respectively.
For the systemic velocity we assume $+22.0$~km\,s$^{-1}$ \citep{duf95}. 

From all CHARA and VLTI interferometric observations, we also derived closure phases (CPs).
Both the $H$- and $K$-band continuum closure phases are consistent with zero on a 
$2\sigma$-level, which leads us to conclude that the brightness distribution does
not show significant indications for deviations from centro-symmetry.
The most constraining CPs have been recorded with MIRC in the $H$-band,
which are shown in Fig.~\ref{fig:bcmicp}.
Our AMBER HR measurements from 2009-12-31 provide us also with
a CP measurement in the Br$\gamma$-line of $\beta$\,CMi, 
while the derived CPs from the 2010-04-23 dataset are rather noisy
and are therefore not included for our model fits.

$\zeta$\,Tau was observed on 2010-01-01 
using AMBER's MR-mode
covering the upper $K$-band (2.12 to 2.46~$\mu$m).
The assumed systemic velocity for $\zeta$\,Tau is
$+21.8$~km\,s$^{-1}$ \citep{duf95}.
In the $\zeta$\,Tau data, we detected not only the Br$\gamma$
transition (7-4), but also hydrogen Pfund transitions 
(Fig.~\ref{fig:ztauphotocenter}, 1st row),
including clear detection of Pf14 (19-5, 2.4477~$\mu$m) 
to Pf22 (27-5, 2.3591~$\mu$m).
Higher Pfund transitions are also present in the spectrum, 
but cannot be clearly separated due to the wide, double-peaked profile
of the individual lines.
For $\zeta$\,Tau, the derived CPs are too noisy to 
provide additional information and are therefore not included in our
further analysis.

Details about the observational setup for all interferometric observations 
are listed in Tab.~\ref{tab:obslog}.  
Each observation on a science star was accompanied by observations on 
interferometric calibrators, allowing us to monitor and correct for the 
atmospheric and instrumental transfer function.

\section{Photocenter analysis}
\label{sec:photocenter}

\begin{figure*}[tbhp]
  \centering
  \includegraphics[angle=0,scale=0.6]{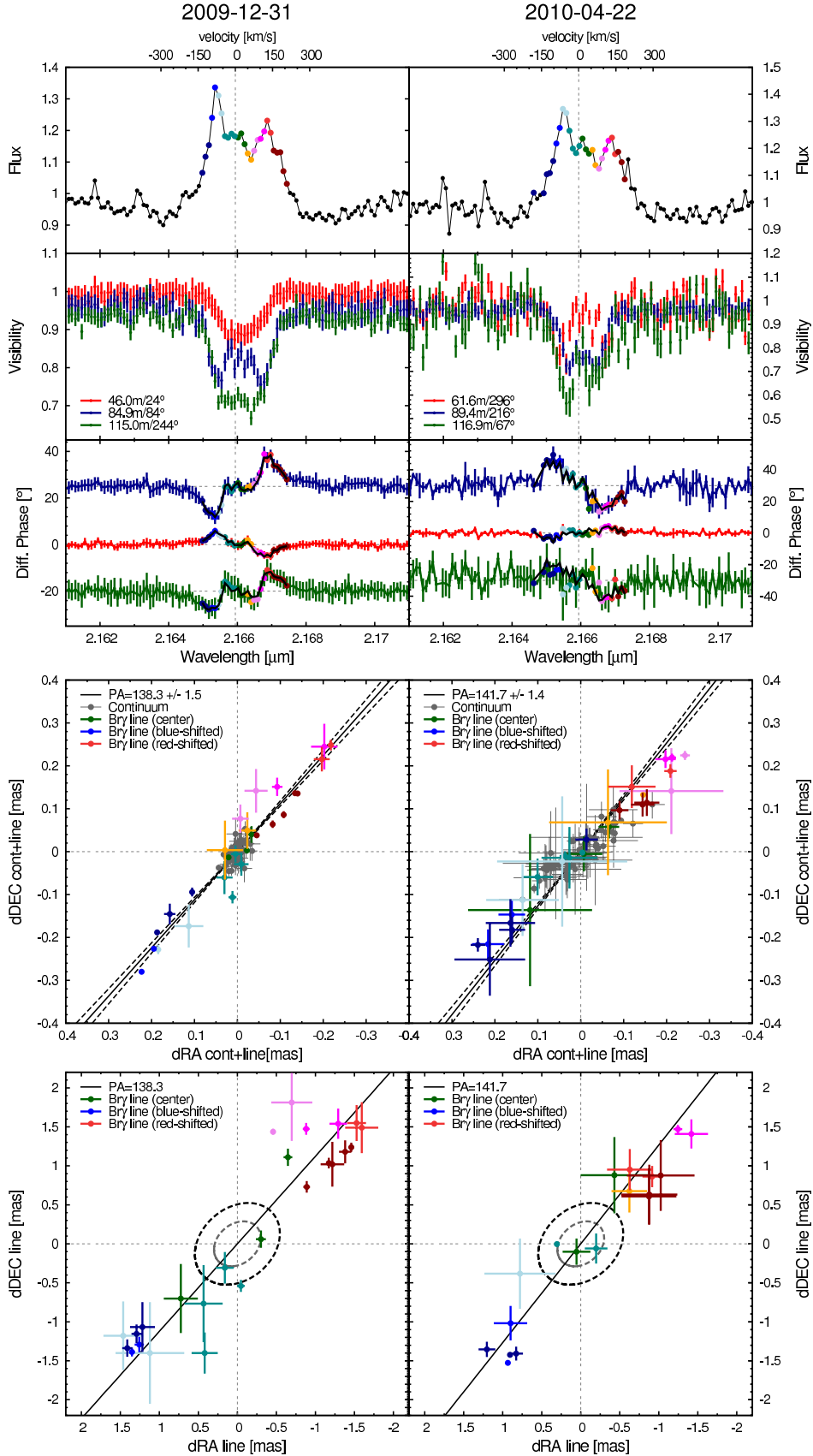}
  \caption{
    \footnotesize
    {\it Upper panel:} 
    VLTI/AMBER spectra {\it (1st row)}, visibilities {\it (2nd row)}, and DPs {\it (3rd row)} measured on 
    $\beta$\,CMi for the epochs 2009-12-31 {\it (left)} and 2010-04-22 {\it (right)}.
    {\it Middle panel:}
    From the measured DPs, we derive for each spectral channel the 2-D photocenter displacement vector 
    (including continuum and line contributions; East is plotted left and North is up).
    The flux and DP data points as well as the derived photocenter vectors have been
    color-coded based on the Doppler velocity 
    (see the upper panel to relate each color to a wavelength).
    The DPs corresponding to the astrometric solutions are shown in the 
    middle panel as solid black lines.
    {\it Bottom panel:} 
    Using the procedure outlined in Sect.~\ref{sec:photocenter}, we corrected
    for the continuum contributions in the line spectral channels,
    revealing the photocenter displacement corresponding to the line emission only.
    Besides the determined photocenter offsets, for comparison we also show
    the size of the $H$-band continuum-emitting disk
    (black ellipse, FWHM Gaussian, Sect.~\ref{sec:bcmicont}) and of the 
    stellar photosphere (grey ellipse, Sect.~\ref{sec:stellarparameters}).
  }
  \label{fig:bcmiphotocenter}
\end{figure*}

\begin{figure*}[tbhp]
  \centering
  \includegraphics[angle=0,scale=0.6]{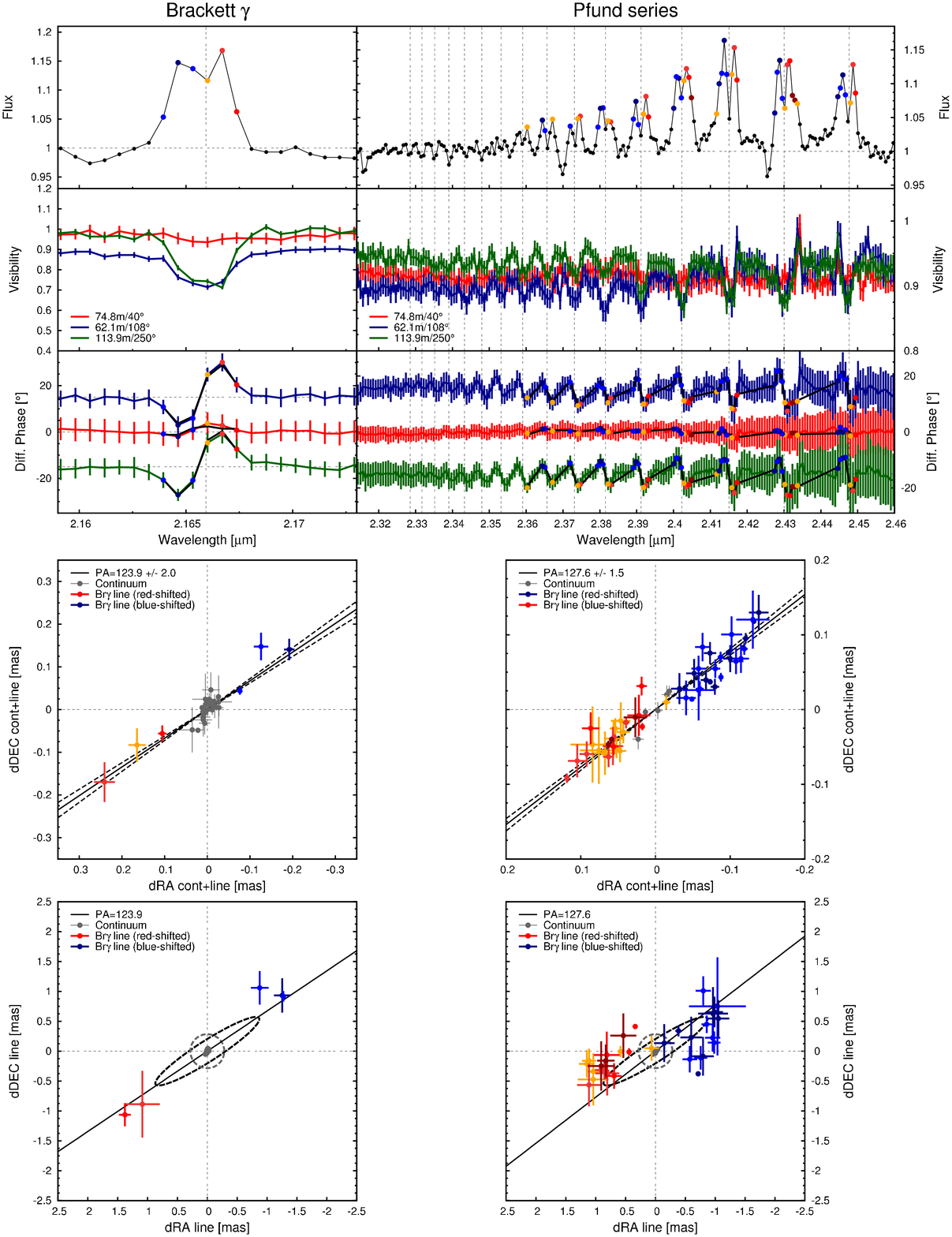}
  \caption{
    \footnotesize
    {\it Upper panel:} 
    VLTI/AMBER spectra {\it (1st row)}, visibilities {\it (2nd row)}, and DPs {\it (3rd row)}
    measured on $\zeta$\,Tau for the epoch 2010-01-01
    in the Br$\gamma$ {\it (left)} and Pf14-22 transitions {\it (right)}.
    {\it Middle/bottom panel:}
    Photocenter displacement vectors derived from the measured DPs (including line and continuum emission, {\it middle panel})
    and the continuum-corrected DPs (tracing the line emission only, {\it bottom panel}).
    For comparison, we also show the size of the $H$-band continuum-emitting disk,
    as determined by \citet{sch10} for epoch 2009-11-10
    (black ellipse, FWHM Gaussian) and of the stellar photosphere (grey ellipse).
  }
  \label{fig:ztauphotocenter}
\end{figure*}

Differential phases provide unique information about small-scale (sub-mas) photocenter
displacements between the blue- and red-shifted line wings. These displacements 
provide a very sensitive measure of the gas kinematics on scales of a few stellar radii.
In a first analysis step, we reconstruct the on-sky 2-D photocenter displacement 
from the measured DPs by solving the system of linear equations
\begin{equation}
  \vec{p} = -\frac{\phi_{i}}{2\pi} \cdot \frac{\lambda}{\vec{B_{i}}},
  \label{eq:photocenter}
\end{equation}
where $\phi_{i}$ is the differential phase measured on baseline $i$,
$\vec{B_{i}}$ is the corresponding baseline vector, 
and $\lambda$ is the central wavelength \citep{leb09}.
The derived photocenter plots for $\beta$\,CMi and $\zeta$\,Tau are shown
in Figs.~\ref{fig:bcmiphotocenter} and \ref{fig:ztauphotocenter} ({\it middle panel}), 
respectively, and clearly reveal rotation-dominated velocity fields for both objects, 
as indicated by the linear alignment of the photocenter vectors for the different gas
velocities and the opposite sign of the photocenter displacement for 
the blue- and red-shifted emission.
In order to associate the DP with the on-sky orientation, it is necessary
to calibrate the sign of the DP measurements.  
For this purpose, we reprocessed the $\zeta$\,Tau VLTI/AMBER data set presented by \citet{ste09} 
and calibrated our DP sign in order to match the published on-sky orientation.

For $\beta$\,CMi, we determine the position angles $\theta$ for the disk rotation plane at the two epochs to be
$138.3 \pm 1.5^{\circ}$ (2009-12-31) and $141.7 \pm 1.4^{\circ}$ (2010-04-22).
The photocenter vectors corresponding to our highest-SNR observation (2009-12-31) 
show an interesting arc-like structure in the red-shifted line wing 
(Fig.~\ref{fig:bcmiphotocenter}, {\it bottom left}),
where the photocenter vectors corresponding to low gas velocities 
are above the derived disk plane, while the high velocities are 
displaced in the opposite direction.
Although the significance of this pattern is still only marginal in our data, 
we speculate that this pattern might result from opacity effects, with the 
more distant parts of the disk appearing fainter than the disk parts facing the observer.
Such an obscuration screen would displace the photocenter perpendicular to the disk plane,
where the amplitude of the displacement is stronger for lower gas velocities, 
since the low-velocity emission is distributed over a more extended region.
Accordingly, the displacement would be strongest at zero velocities, and then symmetrically 
decrease towards higher velocities.  
The superposition of this weak displacement (perpendicular to the disk plane) 
with the displacement due to Keplerian rotation (parallel to the disk plane) 
might result in the observed arc-shaped structure, which would also provide a unique tool 
to determine the orientation of the disk in space and the disk rotation sense.
In the case of $\beta$\,CMi, this implies that the north-eastern part of the disk is 
facing towards the observer (based on the displacement of the low-velocity channels in this direction)
and that the disk is in clockwise rotation (based on the location of the red-shifted 
photocenter displacements in the north-western quadrant).

Another intriguing feature in the $\beta$\,CMi data is the $W$-shaped profile, 
which we observe in the wavelength-dependent visibilities and DP on our 
longest interferometric baselines (Fig.~\ref{fig:bcmiphotocenter}, {\it 2nd} and {\it 3rd row}).
Likely, this profile indicates that the visibility function of the line-emitting region passes 
through a visibility null and transits from the first to the second visibility lobe.
Since this effect would reverse the direction of the photocenter vector, 
we manually correct the phase sign in these corresponding spectral channels close to the line center.
For $\zeta$\,Tau, we can determine the rotation axis for the Br$\gamma$ ($123.9 \pm 2.0^{\circ}$)
and the nine Pf14-Pf22 transitions ($127.6 \pm 1.5^{\circ}$) separately and find 
that the line-emitting gas rotates in the same disk plane within the observational uncertainties of $\sim 2^{\circ}$.

Using the aforementioned procedure, it is possible to reliably measure the direction of the
photocenter displacement, while the length of the displacement vector 
is biased by the contributions from the underlying continuum emission.  
In order to remove these contributions from the measured observables ($F$, $V$, $\phi$), 
we apply the method outlined by \citet{wei07} and interpolate the continuum flux ($F_c$) 
and continuum visibility ($F_c$) from the adjacent continuum.
The visibility and DP of the pure line emitting-region ($V_l$, $\phi_l$) are then
given by
\begin{eqnarray}
  | F_l V_l |^2  & = &  | F V |^2 + | F_c V_c |^2 - 2 \cdot F V \cdot F_c V_c \cdot \cos \phi\\
  \sin \phi_l   & = & \sin \phi \frac{| F V |}{| F_l V_l |},
\end{eqnarray}
where $F_l=F-F_c$ denotes the flux contribution from the spectral line.
The continuum-corrected DPs are then used to derive the photocenter displacement of the
pure line-emitting region (Figs.~\ref{fig:bcmiphotocenter} and \ref{fig:ztauphotocenter}, {\it bottom panel}).
Applying this correction will provide the real centroid offset of the line emission,
but also introduce noise from the visibility and flux measurements,
resulting in an increased scatter in the position angle distribution.
Therefore, we decided to measure the position angle of the rotation axis from the uncorrected line+continuum
photocenter displacements (Figs.~\ref{fig:bcmiphotocenter} and \ref{fig:ztauphotocenter}, {\it middle panel}),
while the continuum-corrected photocenter displacements ({\it bottom panel}) will be used
later on to construct a position-velocity diagram and to compare the spatial origin in different line tracers.

\section{Discussion on $\beta$~Canis Minoris}
\label{sec:bcmi}

$\beta$\,CMi (HR\,2845) is a relatively quiet 
B8V-type classical Be star located at a distance of $52.2^{+2.4}_{-2.2}$~pc \citep{tyc05}.
Ground-based photometric monitoring has provided no clear indications for 
significant variability \citep{pav97}, 
while there is some marginal evidence for long-term variations in the 
H$\alpha$-profile \citep{pol02b,hes09}.    

In the following, we present a refined model for the photospheric emission
of $\beta$\,CMi taking the near-critical stellar rotation into account (Sect.~\ref{sec:stellarparameters}),
followed by our modeling of the interferometric data in continuum emission (Sect.~\ref{sec:bcmicont}).
Given that the disk rotation signatures in the Br$\gamma$ emission line 
are overlayed on the rotation signatures of the star (in Br$\gamma$ absorption), 
we investigate the influence of the stellar rotation on our measurements
(Sect.~\ref{sec:bcmiphotosphere}).
In the following, we construct a position-velocity diagram (Sect.~\ref{sec:bcmiposvel}) 
and present a full kinematical modeling in Sect.~\ref{sec:bcmimodel}.

\subsection{Constraining the stellar parameters}
\label{sec:stellarparameters}

\begin{figure}[bp]
  \centering
  \includegraphics[angle=90,scale=0.38]{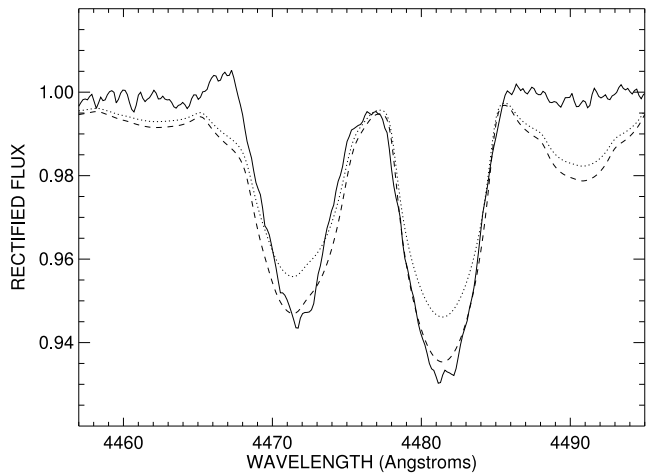}
  \caption{
    \footnotesize
    Spectrum of the \ion{He}{1} 447.1\,nm and \ion{Mg}{2} 448.1\,nm
    line of $\beta$\,CMi (solid line).  The dashed line shows a 
    photospheric model without disk contributions, while the dotted line 
    corresponds to a model where 20\% disk continuum contributions have been added.
  }
  \label{fig:bcmiveiling}
\end{figure}

\begin{table*}[t]
\begin{center}
\caption{Grid of stellar parameters, obtained for $\beta$\,CMi using our fast rotator model (Sect.~\ref{sec:stellarparameters}).}
\label{tab:stargrid}
\begin{tabular}{cccccccccc}
  \tableline\tableline
  \noalign{\smallskip}
  Model       & $i$        & $\omega/\omega_{\rm crit}$ & $M_{\star}$  & $R_{\rm pole}$ & $R_{\rm eq}$ & $T_{\rm pole}$  & $P$     & $L^{\rm app}$   & $T_{\rm eff}^{\rm app}$ \\
              & [\deg]     &                           & [$M_{\sun}$] & [$R_{\sun}$]  & [$R_{\sun}$] & [K]           & [days]  & [$L_{\sun}$]    & [K] \\
  \noalign{\smallskip}
  \tableline
  \noalign{\smallskip}
  A           &  45        & 0.99                      & 3.70       & 3.32         & 4.61         & 13200         & 0.676   & 267           & 11558 \\
  B           &  45        & 0.96                      & 4.11       & 3.03         & 3.94         & 14600         & 0.577   & 343           & 13085 \\
  C           &  45        & 0.93                      & 4.47       & 2.87         & 3.59         & 15650         & 0.527   & 413           & 14235 \\
  D           &  40        & 0.99                      & 3.98       & 2.95         & 4.10         & 14550         & 0.546   & 337           & 12871 \\
  E           &  40        & 0.96                      & 4.49       & 2.74         & 3.56         & 16050         & 0.475   & 435           & 14505 \\
  F           &  40        & 0.93                      & 4.91       & 2.60         & 3.25         & 17300         & 0.433   & 532           & 15847 \\
  G           &  35        & 0.99                      & 4.43       & 2.62         & 3.64         & 16200         & 0.433   & 437           & 14447 \\
  H           &  35        & 0.96                      & 5.03       & 2.44         & 3.17         & 18100         & 0.377   & 590           & 16474 \\
  \noalign{\smallskip}
  \tableline
\end{tabular}
\end{center}
\end{table*}

\begin{figure}[bp]
  \centering
  \includegraphics[angle=0,scale=0.38]{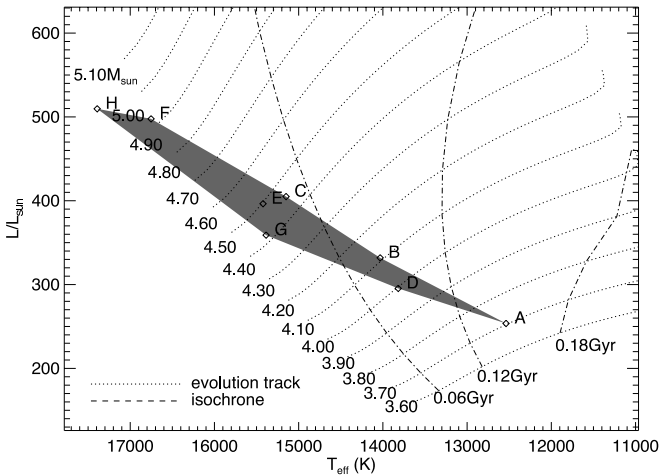}
  \caption{
    \footnotesize
    Hertzsprung-Russell diagram, including the solutions from 
    our small rapid rotator grid for $\beta$\,CMi (Tab.~\ref{tab:stargrid}).
    The numbers on the left side give the stellar mass of the
    corresponding evolution track, while the numbers at the bottom 
    gives the evolutionary age corresponding to the shown isochrones \citet{yi03,dem04}.
    As outlined in Sect.~\ref{sec:stellarparameters}, only models
    A, B, and D satisfy the observational constraints on the 
    stellar effective temperature.
    Furthermore, model~D satisfies best our inclination 
    obtained with interferometry ($38.5\pm1^{\circ}$).
  }
  \label{fig:bcmihrd}
\end{figure}

In order to obtain a model for the photospheric emission of $\beta$\,CMi, 
we searched for a consistent set of stellar parameters,
taking the evidence for near-critical rotation into account \citep{sai07}.
For this purpose, we employed our rapid rotator code \citep{mon07,che11},
which simulates the stellar oblateness and surface temperature distribution using
the modified von Zeipel theorem (gravity darkening coefficient = 0.188), and computes model images 
in the continuum and in the photospheric Br$\gamma$ absorption line.

In order to constrain the stellar parameters, we computed a small parameter grid,
in which we systematically varied the inclination angle $i$
and the fractional angular velocity $\omega/\omega_{\rm crit}$ 
(where $\omega_{\rm crit}$ is the critical angular velocity).
For a given inclination angle and fractional angular velocity, 
we start our iterative process by assuming a stellar mass $M_{\star}$, 
from which we compute the polar radius $R_{\rm pole}$
using the rotation velocity $v \sin i=244 \pm 6$~km\,s$^{-1}$ \citep{yud01}.
The polar temperature $T_{\rm pole}$ is derived from the $V$-band magnitude ($V=2.89$).
Any disk continuum flux in the optical band will tend
to dilute the photospheric spectrum causing absorption
lines to appear weaker than expected.  We checked for
such line reduction in blue spectra of $\beta$\,CMi
obtained by \citet{gru07}.  Fig.~\ref{fig:bcmiveiling} 
shows the observed profiles ({\it solid line}) of
\ion{He}{1}~447.1\,nm and \ion{Mg}{2}~448.1\,nm
based upon the average of 11~spectra made between 2005
and 2009 with the Kitt Peak National Observatory Coude
Feed Telescope (spectral resolution $R=12,500$).
The observed spectrum is compared with a theoretical flux
spectrum ({\it dashed line}) from the grid of LTE models
by \citet{rod05}.  We adopted average parameters for the visible hemisphere 
of the star of $T_{\rm eff} = 11,800$~K, $\log g = 3.8$, and 
$V \sin i = 230$\,km~s$^{-1}$ \citep{fre05}, and the model
spectrum was convolved with a simple rotational broadening
function for a linear limb darkening coefficient of
$\epsilon = 0.42$ \citep{wad85}.   Note that
this approach assumes a spherical star, ignores gravity
darkening, and neglects changes in the local intensity spectrum
with orientation, but these simplifications are acceptable
for our purpose of checking for systematic line depth
differences.  We see that the match of the \ion{Mg}{2} $\lambda 4481$
profile is satisfactory, but the wings of the \ion{He}{1} $\lambda 4471$
line appear too high, probably due to low level emission in
this transition.  Adding a $20\%$ disk flux contribution and
renormalizing the spectrum yields a diluted version ({\it dotted line}),
which appears to be much too weak compared with the observed spectrum.
This comparison suggests that the disk continuum emission in
the blue range is negligible, and thus we will ignore any
disk contribution to the optical flux in the following discussion.

With the derived polar radius and bolometric luminosity, we are able to locate the 
stellar position on the HR diagram, after correcting for the rotational effect \citep{che11}. 
The stellar mass from the HR diagram is used as initial value for the next iteration step,
until convergence between the assumed mass and the mass estimated from the HR diagram is reached.

For each parameter combination, our model provides the rotation period, apparent luminosity $L^{\rm app}$, 
and apparent effective temperature $T_{\rm eff}^{\rm app}$, as listed in Tab.~\ref{tab:stargrid}.
Comparing the model effective temperature and luminosity with the observational
constraints for $\beta$\,CMi \citep[$T_{\rm eff}=12,050$~K; $L=195\pm60~L_{\sun}$;][]{sai07},
we find good agreement for models A, B, and D.
Due to its consistency with the inclination and effective temperature
(Sects.~\ref{sec:bcmicont} and \ref{sec:bcmimodel}), we favor model~D, suggesting
$i=40^{\circ}$, $\omega/\omega_{\rm crit}=0.99$, $T_{\rm pole}=14,550$, $v \sin i=244$~km\,s$^{-1}$, $T_{\rm eff}^{\rm app}=12\,871$~K, 
and a polar and equatorial radius of $R_{\rm pole}=2.95~R_{\sun}=0.26$~mas and $R_{\rm eq}=4.10~R_{\sun}=0.36$~mas, respectively.
We plot the values corresponding to our model grid in the HR diagram shown in Fig.~\ref{fig:bcmihrd}
and compare it to the evolutionary tracks from \citet{yi03} and \citet{dem04}, 
yielding an evolutionary age of $\sim 0.1-0.15$~Gyr.

Obviously, an important input parameter for our modeling procedure
is the rotation velocity $v \sin i$, for which we use an average value from the literature. 
\citet{tow04} argued that most $v \sin i$ measurements on Be stars
might systematically underestimate the true projected rotation value due to the effect 
of gravity darkening.  In order to test this scenario, we have artificially increased the 
measured $v \sin i$-value by 10\% and repeated our modeling procedure.  We find that the increase
in $v \sin i$ results in a significant increase in the stellar mass, apparent effective temperature, 
and apparent luminosity, making the model prediction much less consistent with the observed values.
Therefore, we suggest that in the case of $\beta$\,CMi, the $v \sin i$-measurements are not 
significantly biased by gravity darkening, which is likely a result of the intermediate inclination angle of 
$\beta$\,CMi (the effect would be stronger for an equator-on viewing angle)
and the use of our moderate gravity darkening law ($T_{\rm eff} \propto g_{\rm eff}^{0.188}$, \citealt{che11}),
which is likely more relatistic than the classical von~Zeipel darkening coefficient 
($T_{\rm eff} \propto g_{\rm eff}^{0.25}$) adopted by \citet{tow04}.
Clearly, the accurate determination of $v \sin i$ remains a fundamental problem for constraining the 
stellar parameters of near-critical rotating stars.  Therefore, we are currently working on incorporating the 
computation of model spectra in our modelling procedure, which will then be fitted to observed spectra
together with the photometric and interferometric constraints (Che et al., in prep.).

The modeling provides spectra as well as model images
in the continuum emission and in hydrogen photospheric 
absorption lines, which we will use as a representation of the
stellar brightness distribution for our modeling 
of the disk in the following sections.

\subsection{Continuum disk geometry}
\label{sec:bcmicont}

\begin{figure*}[htbp]
  \centering
  $\begin{array}{c@{\hspace{3mm}}c}
    \includegraphics[angle=0,scale=0.8]{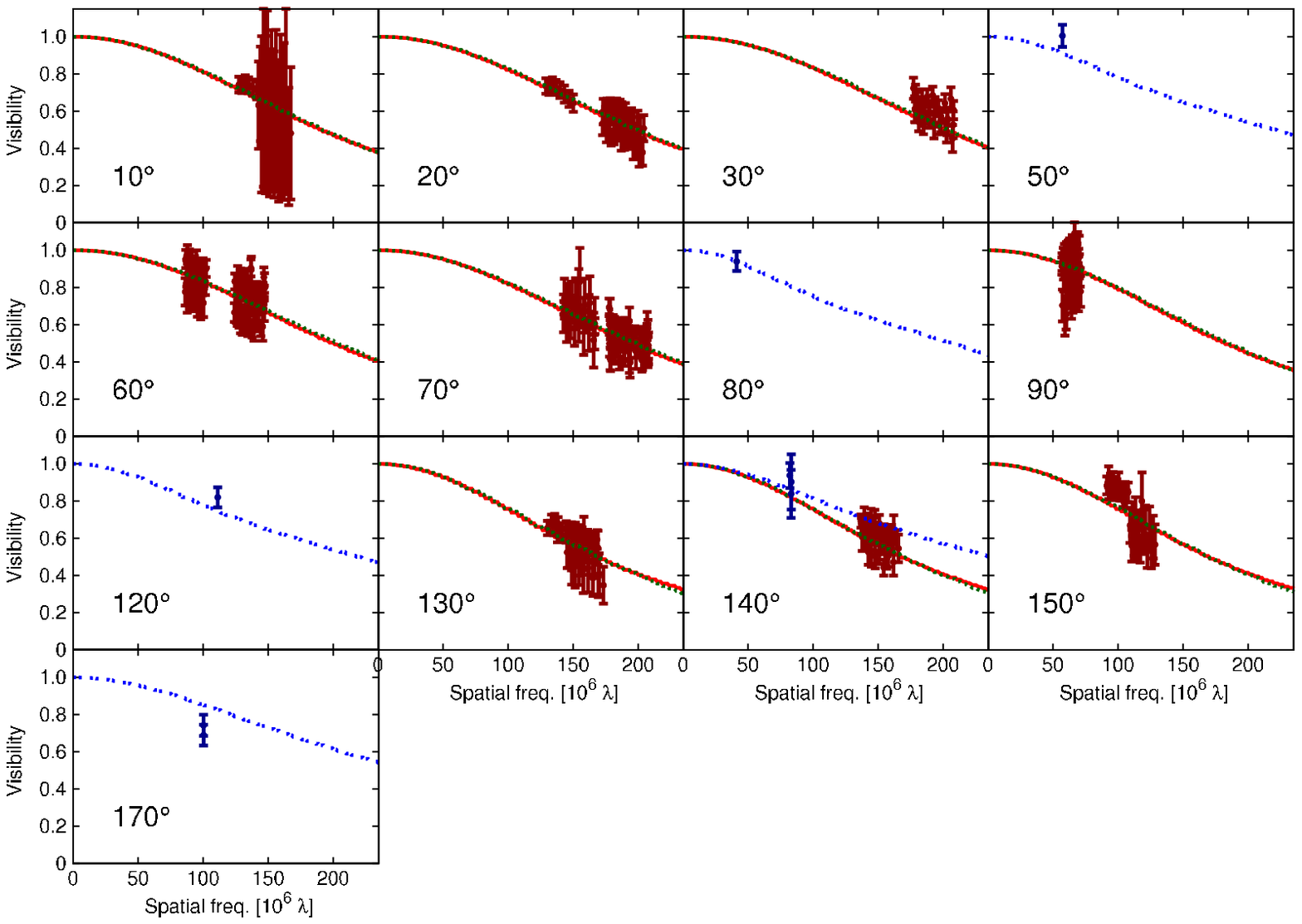} & 
    \begin{minipage}{50mm}
      \vspace{-85mm}
      \includegraphics[angle=0,scale=0.25]{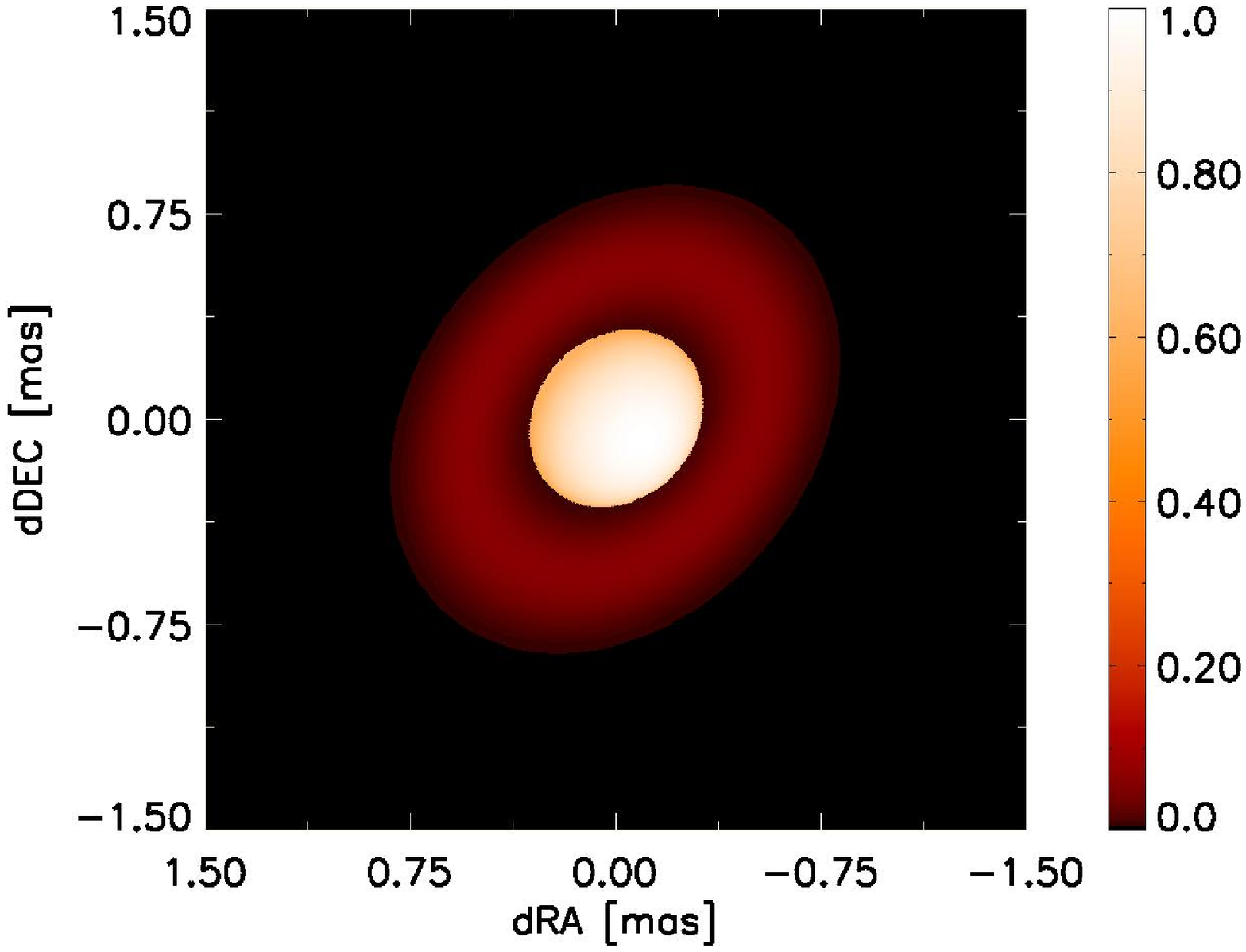} \\
      \includegraphics[angle=0,scale=0.25]{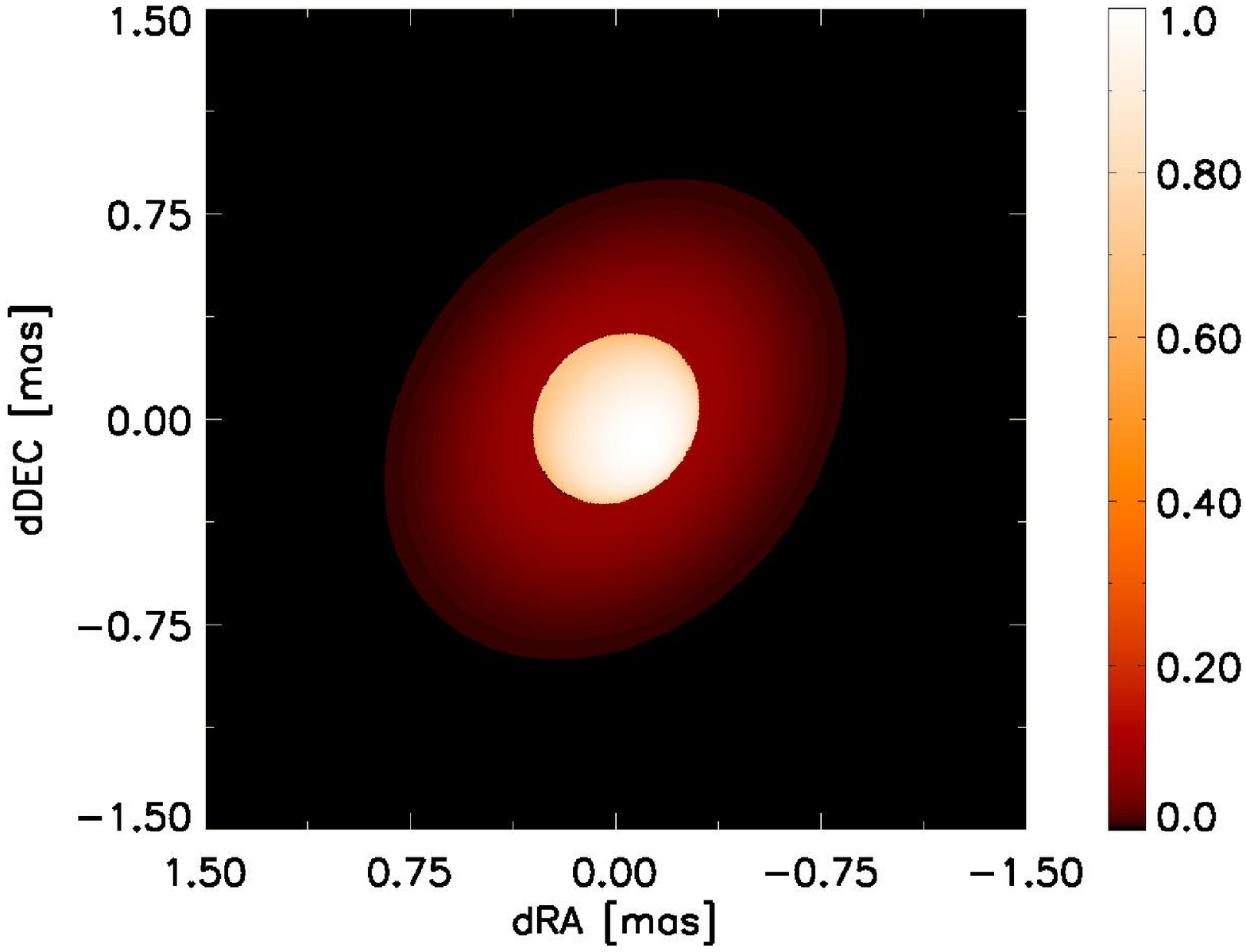} \\
    \end{minipage} \\
  \end{array}$
  \caption{
    \footnotesize
    {\it Left:} Continuum visibilities measured on $\beta$\,CMi 
    with CHARA/MIRC in the $H$-band (red points)
    and CHARA/CLIMB in the $K$-band (blue points).
    In order to show the position-angle dependence of the visibility function, 
    we have binned the data by the projected baseline PA
    (number in lower left corner of each panel).
    For the $H$-band, the best-fit ring model (red line) and Gaussian model (green dotted line) are shown,
    with both models yielding nearly indistinguishable visibility profiles.
    For $K$-band, we show the best-fit Gaussian model (blue dashed line; see Sect.~\ref{sec:bcmicont}).
    {\it Right:} Intensity distribution corresponding to our best-fit ring {\it (top)}
    and Gaussian model {\it (bottom)}.
  }
  \label{fig:bcmicont}
\end{figure*}

\begin{sidewaystable}
\begin{center}
\footnotesize
\caption{Model fitting results for the disk of $\beta$\,CMi.}
\label{tab:results}
\begin{tabular}{lcc|cc|c|cc}
  \tableline\tableline
  \noalign{\smallskip}
                                   &                        &             & \multicolumn{2}{c|}{Continuum}                            & Continuum                & \multicolumn{2}{c}{Line} \\
                                   &                        &             & \multicolumn{2}{c|}{$H$-band}                             & $K$-band                 & \multicolumn{2}{c}{Brackett $\gamma$} \\
                                   &                        &             & (Ring model)                & (Gaussian model)            & (Gaussian model)         & (Keplerian rot.)  & (Free rotation law) \\
  \noalign{\smallskip}
  \tableline
  \noalign{\smallskip}
  Flux ratio                       & $F_{\rm disk}/F_{\star}$ &             & $0.16^{+0.07}_{-0.03}$         & $0.18^{+0.05}_{-0.03}$        & $0.20^{+0.15}_{-0.15}$     & --                 & -- \\
  Major axis size                  & $a$                    & [mas]       & $1.19^{+0.41}_{-0.24}\,^{(b)}$ & $0.33^{+0.18}_{-0.14}\,^{(a)}$ & $0.33^{(a),(c)}$          & --                 & -- \\
  Position angle$^{(e)}$            & $\theta$               & [$\deg$]    & $138.2^{+3.1}_{-7.8}$         & $139.5^{+4.4}_{-6.3}$         & $139.5^{(c)}$            & $136.8^{(c)}$       & $136.8^{(c)}$ \\
  Inclination                      & $i$                    & [$\deg$]    & $39.9^{+7.2}_{-6.4}$          & $40.1^{+5.9}_{-9.2}$          & $40.1^{(c)}$             & $38.5\pm1$          & $38.5^{(c)}$ \\ 
  Outer disk radius                & $R_{\rm out}$           & [mas]       & --                          & --                           & --                      & $5.8\pm0.2$         & $5.8^{(c)}$ \\
  Stellar mass                     & $M_{\star}$             & [$M_{\sun}$] & --                          & --                           & --                      & $3.5\pm0.2$         & $3.5^{(c)}$ \\
  Rotation law index               & $\beta$                &             & --                          & --                           & --                      & $-0.5^{(c)}$         & $-0.5\pm0.1$ \\
  Rotation velocity                & $f_{\rm kep}(1\,\mathrm{AU})$ &       & --                          & --                           & --                      & $1.0^{(c)}$          & $1.0\pm 0.1$ \\
  Radial intensity index           & $q$                    &             & --                          & --                           & --                      & $-1.6\pm0.2$        & $-1.6^{(c)}$\\
  \noalign{\smallskip}
  \tableline
                                   & $\chi^2_r$             &             & 1.77                        & 1.77                         & 1.17                    & 1.66                  & 1.66 \\
  \noalign{\smallskip}
  \tableline
\end{tabular}
\tablecomments{
  $(a)$~For the Gaussian model, we give the half-width half maximum (HWHM) of the half-Gaussian intensity profile, measured along the major axis.
  $(b)$~For the ring models, we give the ring major axis diameter.
  $(c)$~This parameter has been kept fixed throughout the modeling procedure (see Sects.~\ref{sec:bcmicont} and \ref{sec:bcmimodel} for details).
  $(d)$~An inclination angle of $i=0$ denotes a face-on disk orientation.
  $(e)$~Throughout this paper, all position angles are measured along the disk major axis and East of North.
}
\end{center}
\end{sidewaystable}

With projected baseline lengths up to $B=314$~m, our CHARA/MIRC interferometric 
observations allow us to constrain the disk geometry in eight spectral channels in the $H$-band
with an effective resolution of $\lambda/2B=0.5$~mas.
A commonly applied modeling approach is to approximate the disk emission with a Gaussian intensity profile,
which is superposed on a uniform disk representing the stellar surface \citep[e.g.][]{tyc05,mei09,pot10,sch10}.
A caveat of this approach is that a significant part of the disk emission is concentrated 
at radii $r<R_{\star}$, and therefore wrongly attributed to the disk instead of the stellar flux.
Accordingly, these models will systematically overestimate the disk emission with respect to
the stellar emission ($F_{\rm disk}/F_{\star}$), resulting in inconsistencies with SED fitting results.
In order to avoid these biases, we employ an elliptical Gaussian model, where the radial intensity profile
at $r>R_{\star}$ is given by a half-Gaussian.
resulting in a proper estimation of the $F_{\rm disk}/F_{\star}$ ratio.
As alternative model, we considered an inclined ring model, where the radial intensity profile 
is given by a Gaussian centered at radius $a$ from the star, with a fixed fractional width of 25\% 
(see \citealt{mon06a} for details).
Based on radiative transfer simultations of the brightness distribution in classical Be star disks \citep[e.g.][]{wat86},
we consider the Gaussian model a better representation of the expected emission profile in a Be disk, while
the ring model is more suited for comparison with results from the literature \citep[e.g.][]{mei06}.

For all models, the stellar photosphere is represented by the aforementioned rapid rotator model (Sect.~\ref{sec:stellarparameters})
and the position angle $\theta$ of the stellar equator and the disk major axis are aligned.
Besides the disk major axis $a$ (ring major radius or Gaussian half-width half maximum), the PA $\theta$, 
and the inclination $i$, we also treat the flux ratio between the stellar and disk component $(F_{\rm disk}/F_{\star})_{H}$ as a free parameter.
The model was fitted to the MIRC squared visibilities, closure phases, and triple amplitudes using a 
least-square fitting procedure, resulting in the best-fit model shown in Fig.~\ref{fig:bcmicont}.
The best-fit model parameters are displayed in Tab.~\ref{tab:results}, including
1$\sigma$-errors, which we have determined using a boot strapping procedure.
We also tested whether the agreement can be improved using a skewed ring model \citep{mon06a}, 
but find that introducing any disk asymmetry only marginally improves the fit.
Our model includes minor asymmetries due to the brightened polar region in our 
rapid rotator model, which results in small closure phases ($\lesssim 1.2^{\circ}$),
consistent with the measurement (Fig.~\ref{fig:bcmicp}).

The determined disk position angles of $139.2^{\circ}$ (ring model) and $139.3^{\circ}$ (Gaussian model) 
are in excellent agreement with the gas disk rotation axis determined 
with our VLTI/AMBER photocenter analysis ($140.0\pm 1.7^{\circ}$).

Using the detailed information about the $H$-band geometry obtained
with our extensive MIRC data set, we then fitted our Gaussian model to the CLIMB data
to determine the $K$-band continuum geometry.
Given the lower amount of observational constraints for the $K$-band, 
we treated only the disk-to-star flux ratio $(F_{\rm disk}/F_{\star})_{K}$ as a free parameters 
and kept the remaining parameters fixed.
The resulting best-fit visibility curves are shown in Fig.~\ref{fig:bcmicont}.
With $(F_{\rm disk}/F_{\star})_{K}=0.25^{+0.09}_{-0.08}$,
we do not find any significant deviations between the $K$-band and $H$-band flux ratio.

\subsection{Investigating the differential phase signatures of 
  the stellar rotation in the Br$\gamma$ absorption line}
\label{sec:bcmiphotosphere}

\begin{figure*}[t]
  \centering
  $\begin{array}{c}
    \includegraphics[angle=0,scale=0.85]{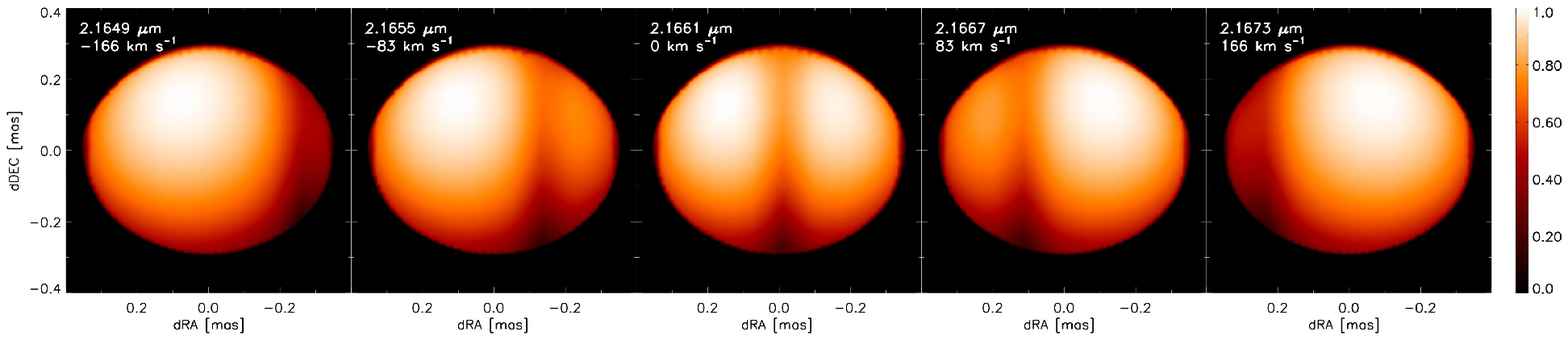}\\[3mm]
    \includegraphics[angle=0,scale=0.85]{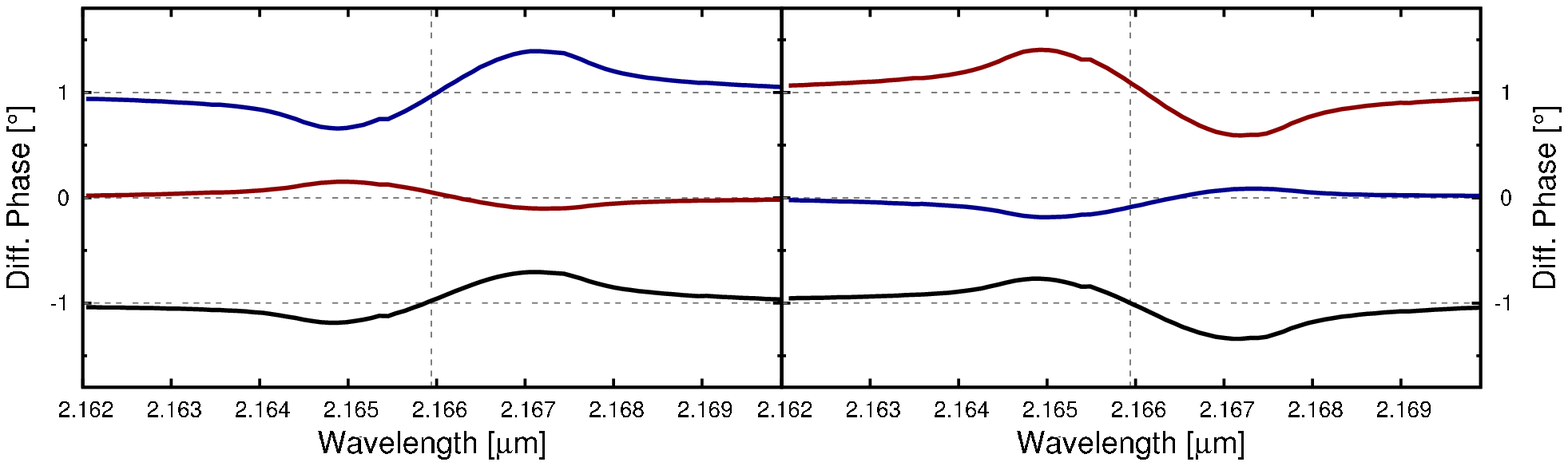}
  \end{array}$
  \caption{
    \footnotesize
    {\it Top:} Model images from our rapid rotator model for $\beta$\,CMi for some representative wavelengths (Sect.~\ref{sec:bcmiphotosphere}).
    {\it Bottom:} Model differential phases (including the stellar photosphere only) computed for the baselines 
    employed by our $\beta$\,CMi VLTI measurements from 2009-12-31 {\it (left)} and 2010-04-23 {\it (right)}.
  }
  \label{fig:bcmiphotosphere}
\end{figure*}

With the currently achievable accuracy, DP measurements can already
reveal photocenter displacements two to three orders smaller than the formal 
angular resolution ($\lambda/2B$).
For instance, our 2009-12-31 AMBER observations exhibit a DP accuracy of $\sim 0.8^{\circ}$ 
(standard deviation over all continuum spectral channels), corresponding to a photocenter displacement
of $\sim 8$ micro-arcsecond (0.008~mas) on the employed 115\,m baseline.
Given that this is much smaller than the equatorial stellar diameter ($\sim 660$ micro-arcsecond), 
it is important to investigate whether the measured DP signatures might also contain contributions
from the photospheric Br$\gamma$ absorption line, which is tracing the stellar rotation and 
is underlying the Br$\gamma$-line emission.

In order to simulate the effect of stellar rotation on our interferometric observables, 
we employ our rapid rotator code and compute the stellar surface brightness 
distribution around the Br$\gamma$-line
with a similar resolution as our AMBER observations (Fig.~\ref{fig:bcmiphotosphere}, {\it top}).
For this, we align the stellar rotation axis with the 
measured rotation axis ($\theta=227.5^{\circ}$, Sect.~\ref{sec:photocenter}).
Br$\gamma$ absorption on one side of the photosphere will shift the
photocenter towards the opposite direction.  
Taking this into account, we adjust the orientation of our model photosphere, so that the 
Br$\gamma$ absorption in the blue-shifted line wing matches the measured direction of 
the photocenter displacement at red-shifted wavelengths.
In addition to the photospheric emission, we include disk emission in our model,
assuming $(F_{\rm disk}/F_{\rm \star})_{K}=0.25$, as determined in Sect.~\ref{sec:bcmicont}.

From the model, we compute the corresponding DPs and find signatures
$\phi \lesssim 0.5^{\circ}$ on all baselines (Fig.~\ref{fig:bcmiphotosphere}, {\it bottom}).  
This result reflects the fact that the fraction of the 
photospheric absorption to the total $K$-band emission is rather small ($<8$\%) and that the
absorption appears as a rather diffuse structure on the extended stellar surface.
We conclude that the photocenter signatures due to stellar rotation (in photospheric absorption lines)
are a minor effect compared to the rotation signatures of the disk (in line emission), 
but nevertheless result in DP signatures comparable to the currently achievable DP accuracy
and are therefore important, in particular for more detailed future studies.

\subsection{Constraints on the disk kinematics from the position-velocity diagram}
\label{sec:bcmiposvel}

\begin{figure*}[tbhp]
  \centering
  \includegraphics[angle=0,scale=0.6]{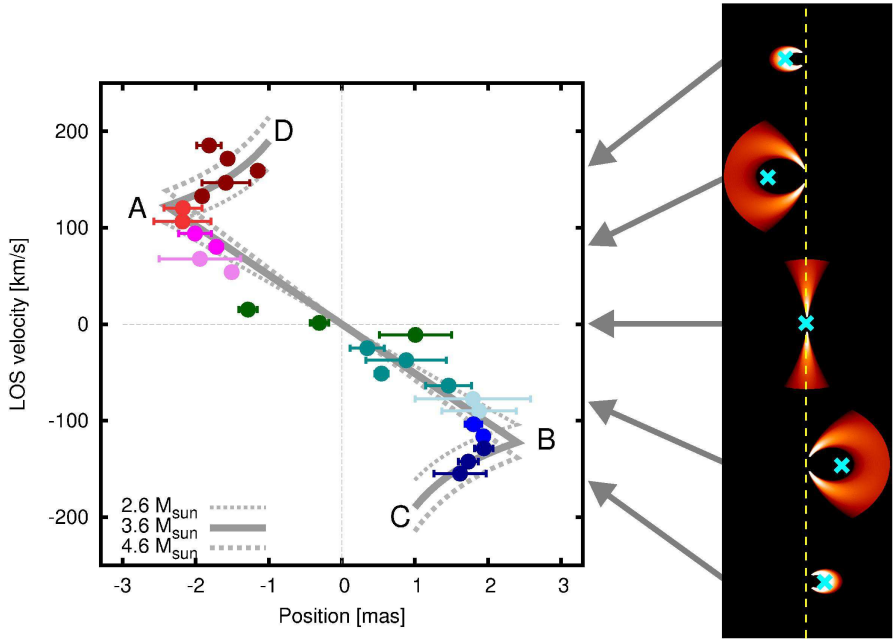}
  \caption{
    \footnotesize
    {\it Left:} Position-velocity diagram, derived from the continuum-corrected photocenter displacement vectors
    in Fig.~\ref{fig:bcmiphotocenter} ({\it bottom, left}).
    The color of the data points matches the color-coding of the wavelength channels in Fig.~\ref{fig:bcmiphotocenter}.
    The model curves show the Keplerian rotation profile for three stellar masses with $R_{\rm in}=0.68$~mas, 
    and $R_{\rm out}=2.4$~mas (Sect.~\ref{sec:bcmiposvel}).
    {\it Right:} In order to illustrate the interpretation of the $p-v$-diagram,
    we show the Br$\gamma$-line model images from our best-fit kinematical model 
    (Sect.~\ref{sec:bcmimodel}, Fig.~\ref{fig:bcmimodel})
    for some representative velocities and mark the corresponding photocenter displacement 
    (corresponding to the position offset plotted on the abscissa in the diagram) 
    in each channel map with a blue cross.
    The yellow vertical line marks the position of the star, which corresponds to the
    photocenter of the continuum emission and the zero position in the $p-v$ diagram.
  }
  \label{fig:bcmiposvel}
\end{figure*}

Position-velocity ($p-v$) diagrams provide a powerful tool for the interpretation 
of a disk velocity field and are commonly employed in radio astronomy to derive the
rotation profile of circumnuclear galactic disks or protostellar disks 
using, for instance, maser \citep[e.g.][]{miy95,pes09} or molecular emission tracers \citep[e.g.][]{sof01,ise07}.
Using our continuum-corrected photocenter displacement measurements, 
we can construct an equivalent diagram from our VLTI interferometric data
by measuring the length of the continuum-corrected photocenter displacement
vectors projected on the disk plane $\theta=140.0^{\circ}$.
Performing this projection on the disk plane also allows us to avoid a
potential bias due to opacity effects, since these effects would move
the photocenter only perpendicular to the disk plane (Sect.~\ref{sec:photocenter}).
The resulting $p-v$ diagram (Fig.~\ref{fig:bcmiposvel}) shows a 
symmetric rotation curve, which we interpret using the simple model 
of a thin Keplerian-rotating disk \citep{wei08}.
The disk extends from an inner ($R_{\rm in}$) to an outer ($R_{\rm out}$) radius
and the line-of-sight (LOS) velocity $v$ of the emission element 
located at radius $r$ and at angle $\vartheta$ in the disk plane is given by
\begin{equation}
  v_{\rm kep}(r,\vartheta) = \sqrt{\frac{GM_{\star}}{r}} \sin \vartheta \cdot \cos i,
\end{equation}
where $\vartheta$ is measured against the LOS and $G$ is the gravitational constant.
Considering only the emission from the outer-most disk annulus (i.e.\ at $R_{\rm out}$ and $\vartheta=-\pi...+\pi$)
will result in a straight line in the $p-v$ diagram (line {\it A-B} in Fig.~\ref{fig:bcmiposvel}).
Adding the emission from the remaining disk annuli ($R_{\rm in}$ to $R_{\rm out}$)
will result in a characteristic ``bowtie''-shaped filled region in the {\it p-v} diagram, 
such as commonly observed in CO imaging observations \citep{sof01,ise07}.
However, astrometric observations (such as radio masers or photocenter displacements)
trace the light barycenter of each annulus, corresponding to the {\it B-C} and {\it A-D} 
curves in Fig.~\ref{fig:bcmiposvel} \citep[for more details see][]{wei08}.
We fit this simple model to the $\beta$\,CMi $p-v$ diagram
assuming a fixed inclination of $i=40^{\circ}$ (as determined in Sect.~\ref{sec:bcmicont}),
which yields best agreement for $M_{\star}=3.6\pm 0.3~M_{\sun}$,
$R_{\rm out}=2.4\pm0.2$~mas, and $R_{\rm in}<1.0$~mas
(solid curve in Fig.~\ref{fig:bcmiposvel}).

\subsection{Detailed modeling of the disk kinematics}
\label{sec:bcmimodel}

\begin{figure*}[t]
  \centering
  $\begin{array}{c}
    \includegraphics[angle=0,scale=0.7]{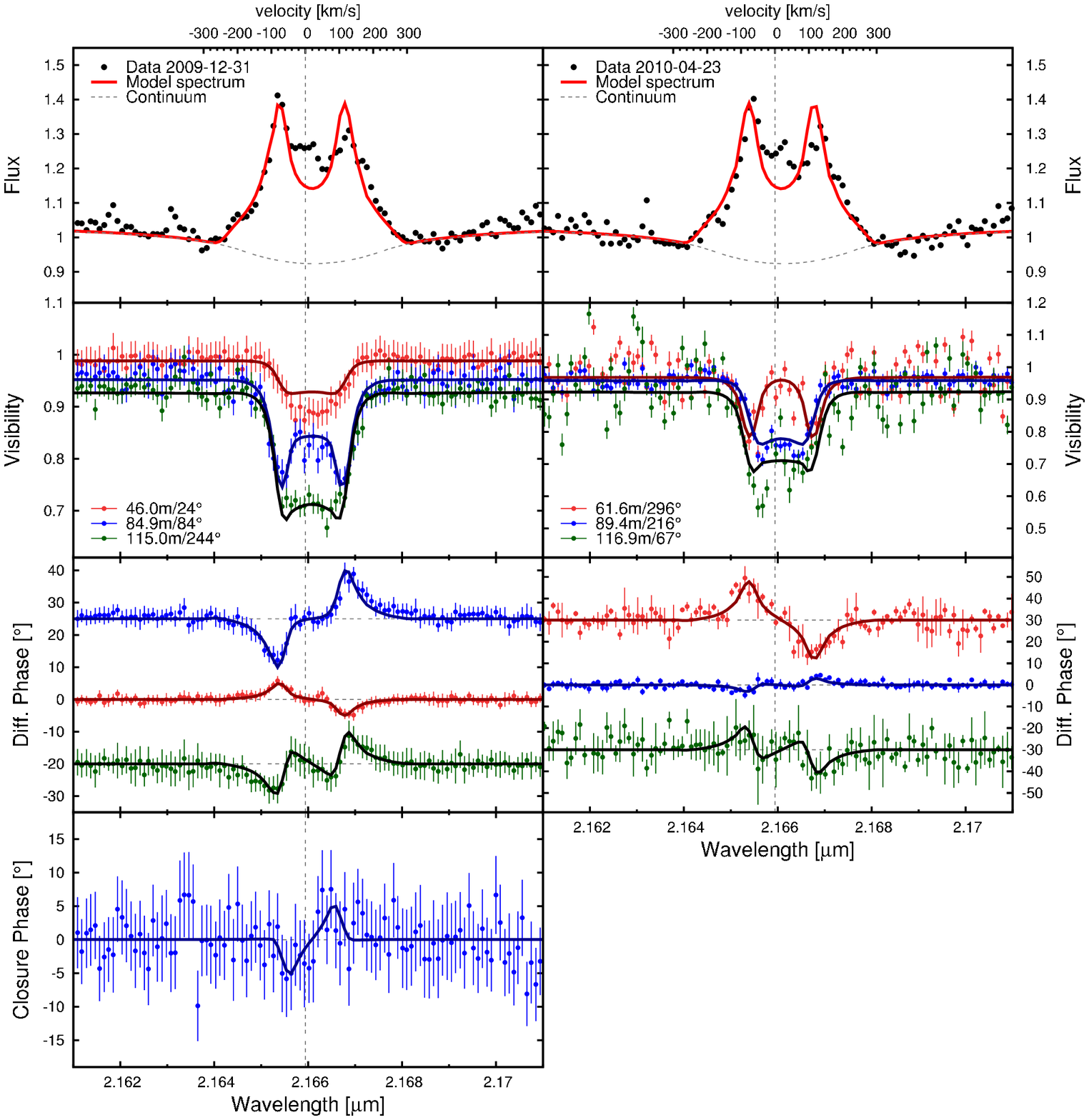}\\[3mm]
    \includegraphics[angle=0,scale=0.9]{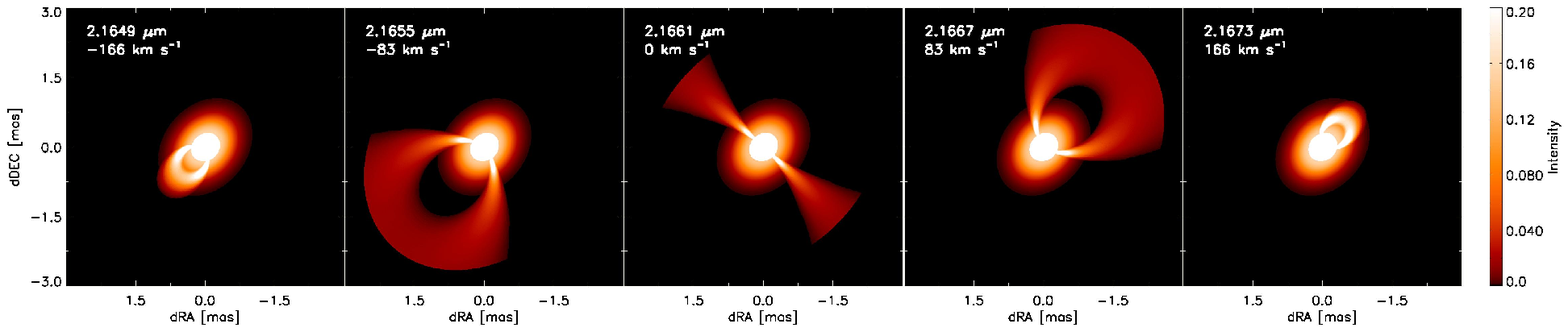}
  \end{array}$
  \caption{
    \footnotesize
    {\it Upper panel:} Comparison of the VLTI/AMBER $\beta$\,CMi spectra {\it (1st row)}, visibilities {\it (2nd row)},
    DPs {\it (3rd row)}, and CPs {\it (4th row)} with our Keplerian disk model (solid lines).
    The three different colors correspond to the three baselines, where the associated projected baseline lengths 
    and PAs are labeled in the 2nd row.
    {\it Lower panel:} To illustrate our kinematical model, we show intensity channel maps 
    for five representative wavelengths.
  }
  \label{fig:bcmimodel}
\end{figure*}

\begin{figure*}[htbp]
  \centering
  \includegraphics[angle=0,scale=0.6]{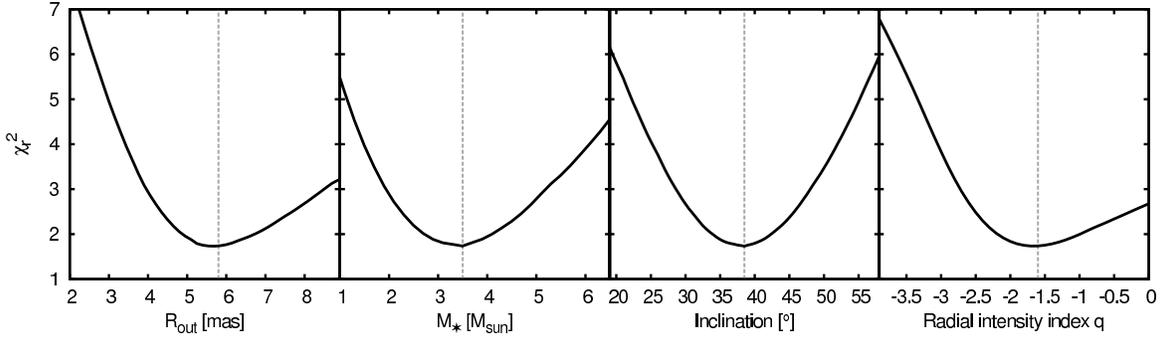}
  \caption{
    \footnotesize
    $\chi^2$-surface derived from our $\beta$\,CMi model grid (Sect.~\ref{sec:bcmimodel}) around the best-fit solution
    assuming a Keplerian velocity field (Tab.~\ref{tab:results}).
  }
  \label{fig:bcmichisq}
\end{figure*}

\begin{figure}[htbp]
  \centering
  \includegraphics[angle=0,scale=0.6]{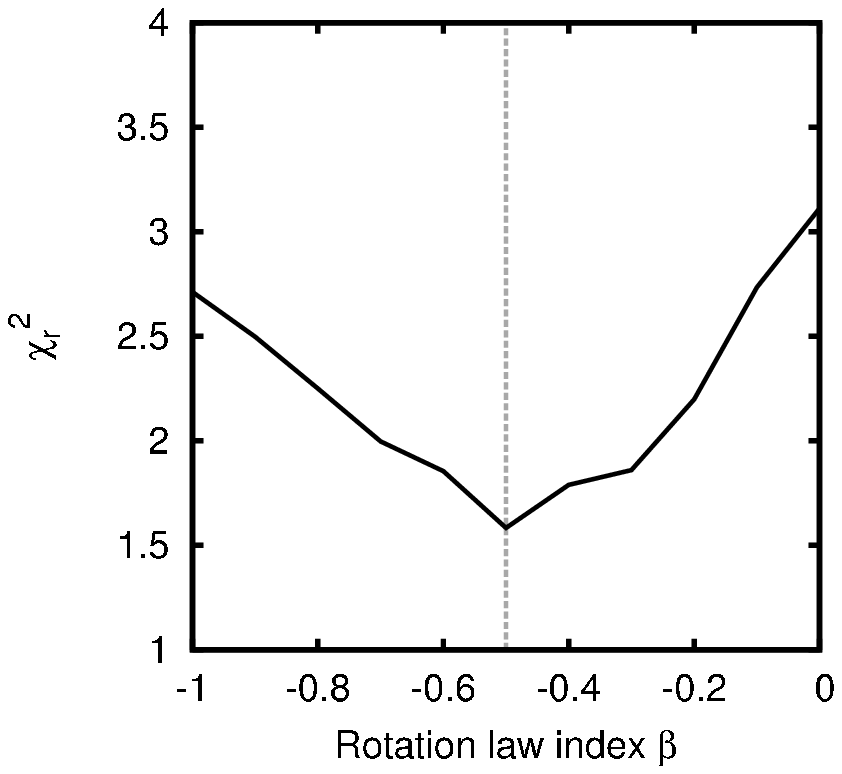}
  \caption{
    \footnotesize
    $\chi^2$-surface for the rotation law index $\beta$
    derived from our $\beta$\,CMi model grid
    including also non-Keplerian velocity fields (Sect.~\ref{sec:bcmimodel}).
  }
  \label{fig:bcmichisq2}
\end{figure}

By reducing the information content to purely astrometric data, the $p-v$ diagram analysis method 
presented in the last section provides a very intuitive and powerful method to constrain the 
gas velocity field in the observed line tracer.
In this section, we will use a more sophisticated model in order to make full use of the
rich information contained in our spectro-interferometric observations,
including spectra, wavelength-differential visibilities, DPs, and CPs.
The Br$\gamma$-emitting gas is assumed to be located in a thin disk plane,
which is justified due to the expected small opening angle of Be star disks
\citep[e.g.][]{bjo93} and the intermediate inclination angle under which
$\beta$\,CMi is observed (Sect.~\ref{sec:bcmicont}).

The gas kinematics is parameterized with a rotation profile 
$|\vec{v}(r)| = f_{\rm kep}(R_{\rm ref}) \cdot (r/R_{\rm ref})^{\beta}$, where 
$\beta=-1$ for rotation with constant angular momentum,
$\beta=-0.5$ for Keplerian rotation, 
$\beta=0$ for constant rotation, and 
$\beta=+1$ for solid body rotation
\citep[e.g.][]{ste96}.
$f_{\rm kep}=|\vec{v}(R_{\rm ref})| / |\vec{v}_{\rm kep}(R_{\rm ref})|$ 
is the orbital velocity at the reference radius $R_{\rm ref} = 1$~AU,
expressed as fraction of the Keplerian velocity $|\vec{v}_{kep}(r)| = (G M_{\star}/r)^{-1/2}$.
In the model, the line emission extends from the 
stellar surface to an outer radius $R_{\rm out}$
with a radial power-law intensity profile $I_{l}(r) \propto r^{q}$.
To include thermal line broadening in our kinematic model, we adopt a
constant radial gas temperature of $0.6 \cdot T_{\rm eff}$ (=7860\,K for $\beta$\,CMi), 
as suggested by the radiative transfer modeling from \citet{car06}.

We include both line and continuum emission in our model and compute 
the interferometric observables for our given VLTI array configurations 
and the covered wavelength channels.
The wavelength-dependent model visibilities and phases 
are then fitted to the VLTI interferometric data using a
reduced $\chi^2_r$ goodness-of-fit estimator
(see \citealt{kra09b} for a definition), including our
flux, visibility, DP, and CP constraints.

For the continuum emission, we assume the geometry 
determined with our CHARA observations (Sect.~\ref{sec:bcmicont})
and use this model to renormalize the AMBER continuum visibilities.
In order to incorporate the underlying photospheric Br$\gamma$ absorption,
we include the photosphere model discussed in Sect.~\ref{sec:bcmiphotosphere},
although, as discussed above, the influence on the differential phase ($\lesssim 0.5^{\circ}$ on all baselines)
is still within the measurement uncertainities.

The disk position angle is fixed to $\theta=140.0^{\circ}$, as determined by our 
model-independent photocenter analysis (Sect.~\ref{sec:photocenter}).
The remaining six parameters in our modeling are the outer disk radius $R_{\rm out}$, 
the inclination $i$, the stellar mass $M_{\star}$, the radial intensity power-law index $q$,
the disk rotation index $\beta$, and the velocity at the reference radius $R_{\rm ref}$ 
(expressed as fraction of the Keplerian velocity, $f_{\rm kep}$).
In a first step, we fix the velocity profile to Keplerian rotation ($\beta=-0.5$, $f_{\rm kep}=1$)
and vary the remaining four parameters systematically on a grid,
yielding the best-fit model shown in Fig.~\ref{fig:bcmimodel}.
The corresponding parameters and uncertainties are listed in column~4 of Tab.~\ref{tab:results}.
In a second step, we test also non-Keplerian velocity fields, 
yielding the best-fit values listed in column~5 (Fig.~\ref{fig:bcmichisq2}).
The resulting $\chi^2$-surfaces are shown in Fig.~\ref{fig:bcmichisq}
and the uncertainties have been derived using the bootstrapping technique.

Our model fits show that the intriging phase inversion observed at the line center
on our longest VLTI baselines (Fig.~\ref{fig:bcmimodel}, {\it 3rd row}) can be explained 
with the phase jumps in the Fourier phase crossing a visibility null.  
These phase jumps appear in the same spectral channels where we measure
visibility minima in the $W$-shaped visibility profile, indicating that the visibility
function of the pure line-emitting geometry transits here from the first to the second visibility lobe
(Fig.~\ref{fig:bcmimodel}, {\it 2nd} and {\it 3rd row}).
At the same time, the continuum emission is only marginally resolved, which results
in the measured composite line+continuum visibility with a rather high contrast of $\sim 0.7$.
The phase jumps and visibility minimums are a basic property of the Fourier transform of 
strongly resolved objects and are reproduced very naturally and without finetuning 
from our kinematical modeling.
Therefore, our results do not support the idea outlined by \citet{ste11}
that these features might indicate secondary dynamical effects or polar mass outflows.

As best-fit value for the radial intensity index, we yield $q=-1.6\pm0.2$.
Isothermal viscous decretion disk models, such as discussed in \citet{bjo05},
predict a radial disk surface density law $\Sigma(r) \propto r^{-2}$, 
corresponding to $q=-2$ in the optically thin case.
More realistic non-LTE disk models suggest a more shallow surface density profile ($\Sigma(r) \propto r^{-2...-1}$)
due to a steep temperature drop in the inner few stellar radii of the disk \citet{car08}.
Therefore, we conclude that our measured intensity profile is well consistent with these models.

The most significant deviation of our simple kinematical model from the data is in the
precise shape of the Br$\gamma$-spectrum.
For instance, the measured spectra show a weak emission component in the line center (Fig.~\ref{fig:bcmimodel}),
which is not reproduced by the model and which might be related either to the presence of
a uniformly distributed low-velocity gas halo or, more likely, to radiative transfer effects.

Another deviation between the model and the observation concerns the weak $V/R$-asymmetry 
which can be observed in our VLTI/AMBER Br$\gamma$ spectra at both epochs (Fig.~\ref{fig:bcmiphotocenter}, {\it top row})
and which is not reproduced by our axialsymmetric kinematical model.
Within the 113~days covered by our AMBER observations, no changes in the $V/R$ asymmetry
could be observed, which is consistent with the conclusion
by \citet{tyc05} and \citet{jon11} that the H$\alpha$ profile does not show significant 
long-term variability.
It is interesting to compare the average photocenter displacement between the blue- and red-shifted line 
emission (2.5~mas) with the characteristic size of the H$\alpha$-emitting region (2.13~mas, \citealt{tyc05}), 
which suggests that Br$\gamma$ emerges from a similar spatial region as H$\alpha$
and a much more extended region than the $H$- and $K$-band continuum emission 
($R_{\rm Br\gamma} \sim R_{\rm H\alpha} > R_{\rm cont}$, Fig.~\ref{fig:bcmiphotocenter}),
in agreement with the prediction from \citet{car10}.
However, a detailed comparison is difficult due to the non-simultaneity of the different observations.

\section{Discussion on $\zeta$\,Tauri}
\label{sec:ztau}

\begin{figure}[bp]
  \centering
  \includegraphics[angle=0,scale=0.65]{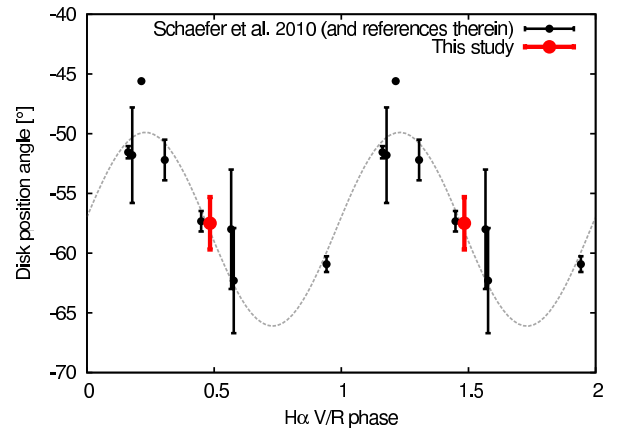}
  \caption{
    \footnotesize
    $\zeta$\,Tau disk position angle, plotted as function of
    the H$\alpha$ $V/R$ phase, including data presented in \citet{sch10}
    and in this study.
    Our new phase measurement (phase 0.484) is in good agreement
    with the disk precession law proposed by \citeauthor{sch10}
  }
  \label{fig:ztauphase}
\end{figure}

$\zeta$\,Tau is a particularly well-studied classical Be star 
which shows a cyclic variability in the flux ratio of the 
violet- and red-shifted wing of the double-peaked 
H$\alpha$ emission line \citep{riv06,ste09}.
These $V/R$ variations exhibit a period of $\sim$1429~days 
and are generally attributed to a global one-armed density oscillation
in a Keplerian (or nearly Keplerian) disk \citep{oka91}.

A kinematic model for the Br$\gamma$-line of some earlier 
AMBER MR observations on $\zeta$\,Tau was presented by 
\citet{ste09} and \citet{car09} and suggested a Keplerian rotation profile.
Our spectro-interferometric observations on $\zeta$\,Tau 
provide a Br$\gamma$ measurement at a new epoch in the H$\alpha$ $V/R$-phase
and cover, for the first time, also the wavelength region around the 
hydrogen Pfund lines.
Besides our Br$\gamma$ and Pfund series data,
we include H$\alpha$-sizes from the literature, 
namely the GI2T photocenter measurement by 
\citet[][$\sim 7~R_{\star}$ or $\sim 1.33$~mas]{vak98}
and the Mark~III and NPOI measurements by \citet{qui94,qui97} and \citet{tyc04}.
It is important to note that the Mark~III and NPOI results were
based on visibility amplitudes instead of differential phases.
Also, a direct comparison with these earlier observations
is complicated by the known V/R variability of $\zeta$\,Tau,
although we note that our measurement at phase 0.484
is reasonably close to the H$\alpha$-measurement
at phase 0.577, presented by \citet{tyc04}.
We convert the Gaussian FWHM derived by these studies
to photocenter displacements by computing the 
centroid of the corresponding Gaussian brightness distributions.
We find that Br$\gamma$ emerges from a similar spatial region than H$\alpha$,
but a more extended region than the near-infrared continuum emission,
as already found for $\beta$\,CMi (Sect.~\ref{sec:bcmimodel}).
The Pfund emission originates from intermediate stellocentric radii
($R_{\rm H\alpha} \sim R_{\rm Br\gamma} > R_{\rm Pf} > R_{\rm cont}$; Fig.~\ref{fig:ztauphotocenter}).

\subsection{Signatures of the known one-armed oscillation}

Both in Br$\gamma$ and in the Pf14-Pf22 lines, we detect a double-peaked
line profile and clear rotation signatures in the differential phases.
All line transitions exhibit a photocenter displacement with a stronger amplitude 
in the south-eastern (red-shifted) than in the north-western (blue-shifted) lobe (Fig.~\ref{fig:ztauphotocenter}, {\it middle panel}).
Such an asymmetric displacement is consistent with the presence of a one-armed
oscillation in the disk \citep{ste09,car09}.
CHARA/MIRC observations by \citet{sch10} showed that the 
one-armed oscillation pattern can also be observed as an asymmetry in the
$H$-band continuum emission. Using multi-epoch data, they find that the position
angle of the asymmetry is correlated with the spectroscopic variability and
precesses around the star with the H$\alpha$ $V/R$ period.
Our observation (2010-01-01) adds an additional epoch for this analysis
at the $V/R$ phase 0.484 (assuming maximum phase at JD=2454505.0 
and a period of 1429~days, \citealt{sch10}). 
We find that the spiral density maximum is located towards the south-east
of the central star and the measured PA is consistent with the sinusoidal 
PA modulation proposed by \citet{sch10}, as shown in Fig.~\ref{fig:ztauphase}.

The reported polarization angle for $\zeta$\,Tau 
($32.9-33.1^{\circ}$, \citealt{qui97}; $35 \pm 4^{\circ}$, \citealt{gho99}; $32-33.5^{\circ}$, \citealt{mcd99})
is in excellent agreement with the rotation axis position angle ($35.8 \pm 2.0$) 
found by our spectro-interferometric observations.

\subsection{Probing the disk excitation structure using multi-transition spectro-interferometry}
\label{sec:LTEmodel}

\begin{figure}[bp]
  \centering
  \includegraphics[angle=0,scale=0.65]{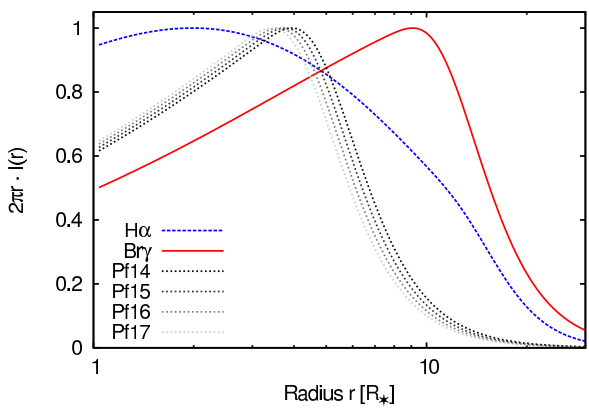}
  \caption{
    \footnotesize
    Radial intensity profile, as computed with our LTE model for 
    different line transitions (Sect.~\ref{sec:LTEmodel}). 
    The intensity has been weighted by the emitting area and
    normalized to the peak intensity.
  }
  \label{fig:ztauLTEmodel}
\end{figure}

\begin{figure}[bp]
  \centering
  \includegraphics[angle=0,scale=0.65]{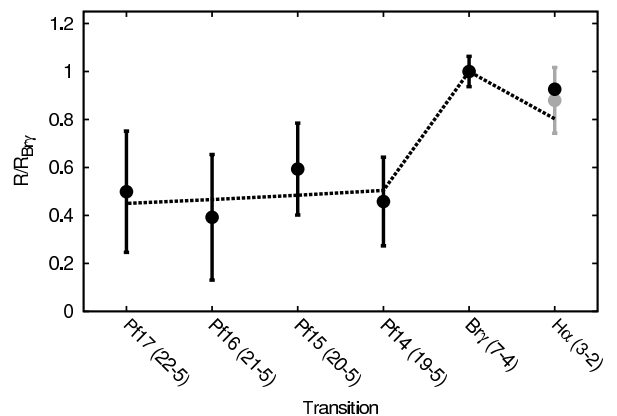}
  \caption{
    \footnotesize
    Origin of the Br$\gamma$ and Pfund line emission from $\zeta$\,Tau,
    as measured by the length of the photocenter displacement vector,
    averaged over all velocity channels for a given line transition
    (the error bars represent the standard deviation in the different velocity channels).
    In addition, we include the H$\alpha$-line photocenter displacement by
    \citet[][black data point]{vak98} and 
    the H$\alpha$ size estimates from \citet{qui94}, \citet{qui97}, \citet{tyc04}, 
    which we converted from Gaussian FWHM to the corresponding photocenters (grey data point).
    The dashed line shows the prediction from our LTE excitation model (Sect.~\ref{sec:LTEmodel}).
  }
  \label{fig:ztausize}
\end{figure}

In order to test whether the measured differences in the stellocentric emitting radius of the
Br$\gamma$ and Pfund transitions are consistent with the expected excitation structure
in the disk, we construct a simple radiative transfer model assuming 
local thermodynamic equilibrium (LTE).  
We assume that the line-emitting material is located in an equatorial disk
which extends outwards from the stellar radius with a constant vertical density per unit volume
and a half-opening angle of $\Theta=5^{\circ}$.
Then, we integrate for each radius $r$ the optical depth $\tau_{\nu}$ in vertical direction, 
assuming hydrogen under LTE conditions and an isothermal temperature distributions.
The number density for the different excitation levels and ionization stages is
computed using the Saha and Boltzmann equation assuming a Gaussian 
line profile with thermal line broadening \citep{wil09}.
The Einstein coefficients for the different hydrogen transitions are estimated 
using the series expansion published by \citet{omi95}.
The radial temperature and surface density profile are parameterized with $T(r) = T_{\rm eff} (r/R_{\star})^{-0.5}$ 
\citep{ste94} and $\Sigma(r) = \Sigma_{0} (r/R_{\star})^{-2}$ with $\Sigma_{0}=2.1$~g\,cm$^{-2}$ \citep{car09}.
For the stellar temperature, equatorial stellar radius, and distance, we assume
$T_{\rm eff}=19\,370$~K, $R_{\star}=7.7~R_{\sun}$, and $d=126$~pc, respectively \citep{car09}.

Using the radiative transfer equation for LTE conditions ($I_{\nu}(r)=B_{\nu}(T) (1-e^{-\tau_{\nu}(r)})$), 
we compute the emitted intensity per unit area as function of radius $r$ in the disk,
where $B_{\nu}$ is the Planck spectrum for temperature $T$ at frequency $\nu$.
From the radial intensity profiles (Fig.~\ref{fig:ztauLTEmodel})
we compute the centroid of the brightness distribution, 
which is then compared to the stellocentric emission radius measured
in the different line transitions.
Given that our model assumes a simplified vertical density structure
and does not include inclination effects, we do not aim to match the
absolute sizes of the emitting region in all line transitions,
but focus instead on the {\it relative} sizes.
For this step, we normalize both the measured and the model 
photocenter offsets relative to Br$\gamma$.
The comparison between the H$\alpha$, Br$\gamma$, and Pf14-17 relative sizes
and our LTE model is shown in Fig.~\ref{fig:ztausize}.
We find that we can reproduce important observational features, in particular that

\begin{enumerate}
\item[\it a)] Br$\gamma$ originates from a similar spatial region in the disk 
as H$\alpha$ ($R_{{\rm Br}\gamma} \sim R_{{\rm H}\alpha}$).
This result is also in agreement with the predictions from
more sophisticated non-LTE radiative transfer computations,
such as made by \citet{car10}.
In order to better characterize the differences between the 
Br$\gamma$ and H$\alpha$-emitting region, contemporaneous observations 
in these two wavelength bands will be required.\\[-5mm]
\item[\it b)] Br$\gamma$ originates from a more extended region
than the Pfund lines ($R_{{\rm Br}\gamma} > R_{\rm Pf}$).
Computing the weighted average of the measurements in the individual transitions, 
we yield $R_{\rm Pf}/R_{\rm {\rm Br}\gamma}=0.48 \pm 0.12$.
For the Pfund lines, we are not aware of earlier predictions obtained
with non-LTE radiative transfer codes.
\end{enumerate}

\vspace{-1mm}
Our observational results confirm the finding from \citet{pot10},
which found $R_{{\rm Br}\gamma} > R_{\rm Pf}$ on the classical Be star 48~Lib,
and consolidates their suggestion that the measured size differences
can already be explained with the expected optical depth differences 
between these line transitions.  We encourage theoreticans and 
modelers to employ their more sophisticated non-LTE radiative transfer codes
in order to test the influence of inclination and non-LTE effects,
although, based on the results from \citet[][e.g.\ Fig.~3]{iwa08}, we expect no 
significant departure from LTE for the inner disk regions 
and the high transitions traced by our observations.
Future multi-transition spectro-interferometric observations 
with improved $uv$-coverage might also measure the precise radial intensity profile 
in a model-independent fashion.
Together with sophisticated radiative transfer simulations,
these observations will reveal the excitation structure 
of the disk and constrain parameters such as the temperature profile
and the vertical disk structure, which are currently difficult to access.

\section{Conclusions}
\label{sec:conclusions}

Using CHARA and VLTI interferometry, our study combined
high angular resolution (with baseline lengths up to 330\,m)
with kinematical information obtained at high spectral dispersion, 
yielding direct observational constraints on the gas distribution, 
excitation structure, and kinematics of the disks around two classical Be stars.
Using a model-independent photocenter analysis method we derived
the disk rotation axis for the prototypical objects $\beta$\,CMi and $\zeta$\,Tau
and spatially and spectrally resolved the disk rotation profile on scales of a few stellar radii.
For both objects, we find that the determined gas rotation plane agrees well 
with the orientation of the continuum-emitting disk as resolved by CHARA,
although there is also clear evidence for substructure in the disk around $\zeta$\,Tau, 
revealing a one-armed oscillation, as indicated by different displacement amplitudes 
in the blue- and red-shifted line wing.

Using our data set on $\beta$\,CMi we constructed a position-velocity diagram,
which can be interpreted using the well-established procedures from radio interferometry,
but probes the milli-arcsecond scale position displacements resulting from the rotating
disk around this Be star.
From our kinematical constraints, we derive the dynamical mass of the central star to $3.5\pm0.2$,
which is in excellent agreement with earlier spectroscopic studies \citep{sai07}.
The inclination of the system is $38.5\pm1^{\circ}$, as determined with
our CHARA continuum and VLTI line observations.
As shown with our detailed kinematical modeling, the rotation law is Keplerian ($\beta=-0.5\pm0.1$) 
and we do not have to include an expanding velocity component in order to explain our data.
Furthermore, our kinematical model allowed us to identify the origin of the phase inversion,
which has now been observed in the differential phases in five out of eight Be stars.
These phase jumps correspond to the transition from the first to the second visibility lobe,
removing the necessity for speculations beyond the canonical star+disk paradigm \citep{ste11}.

For $\zeta$\,Tau, we obtained spectro-interferometric observations covering simultaneously
the Br$\gamma$ and at least nine transitions from the Pfund line series.
For all transitions, we detect a significantly stronger photocenter displacement in the
red-shifted line wing than in the blue-shifted line wing, tracing the one-armed oscillation
which has been deduced for the $\zeta$\,Tau disk before.
Comparing the photocenter displacement in the different line transitions, 
we find that the Pfund, Brackett, and Balmer lines originate from different
stellocentric emitting regions ($R_{\rm cont} < R_{\rm Pf} < R_{{\rm Br}\gamma} \sim R_{\rm H\alpha}$), 
which we can reproduce qualitatively with a simple LTE line radiative transfer model.
More work, including non-LTE radiative transfer modeling, will be required
in order to derive quantitative constraints.

By detecting a purely Keplerian velocity field, our observations are inconsistent with
disk-formation mechanisms incorporating a strong outflowing velocity component,
such as the wind compression scenario, which predicts a strong radial velocity component
comparable to the escape velocity \citep{bjo93}.
On the other hand, our kinematical constraints, as well as the measured
hydrogen line intensity profiles (with a radial power law index $q=-1.6\pm0.2$)
are consistent with the predictions from Keplerian viscous decretion disk models \citep{lee91}.
As shown by \citet{kat83} and discussed in various reviews \citep[e.g.][]{car10}, 
viscous Keplerian disks are also able to produce one-armed density oscillations, 
such as detected for $\zeta$\,Tau.

Considering that our kinematic constraints have been obtained using 
a very limited number of individual measurements 
(2 pointings on $\beta$\,CMi, 1 pointing on $\zeta$\,Tau), 
our study also illustrates the high effectiveness achievable with spectro-interferometry, 
in particular if a very high spectral resolution 
is employed or several line transitions are observed.

\acknowledgments

We thank A.\ Carciofi for helpful discussions on polarization effects in the infrared, 
A.\ M\'{e}rand for validating our spectral line imaging code, 
N.\ Morrison for providing Ritter spectra,
and C.\ Jones for providing us information about the photometric variability of $\beta$\,CMi.
This work was done in part under contract with the California Institute of Technology (Caltech), 
funded by NASA through the Sagan Fellowship Program (SK is a Sagan fellow).
JDM and GHS acknowledge support for this work provided by the National Science Foundation under grants
AST-0707927 and AST-1009080.
The MIRC beam combiner was developed with funding from the University of Michigan.
The CHARA Array is funded by the Georgia State University, by the National Science Foundation 
through grant AST-0908253, by the W.M.\ Keck Foundation, by the NASA Exoplanet Science Institute, 
and the David and Lucile Packard Institute.

{\it Facilities:} \facility{CHARA}, \facility{VLTI}.

\bibliographystyle{apj}
\bibliography{bcmi}

\end{document}